\documentclass[10pt,journal,compsoc]{IEEEtran/IEEEtran}
\usepackage{fullpage}
\usepackage{todonotes}
\usepackage{amssymb}
\usepackage{amsmath}
\usepackage{listings}
\usepackage{color}
\usepackage{hyperref}
\usepackage{graphicx}
\usepackage{caption}
\usepackage{subcaption}

\definecolor{lightgrey}{cmyk}{0.05,0.05,0.05,0}

\newcommand{\figref}[1]{Figure~\ref{#1}}
\newcommand{\secref}[1]{Section~\ref{#1}}
\newcommand{\lstref}[1]{Listing~\ref{#1}}

\begin{document}

\title{Teaching Python programming with automatic assessment and
  feedback provision}

\author{Hans Fangohr, Neil O'Brien, \\
Faculty of Engineering and the Environment\\
University of Southampton \\~\\
Anil Prabhakar, Arti Kashyap \\
IIT Madras, IIT Mandi}
\date{ \today }
\maketitle


\begin{abstract}
We describe a method of automatic feedback provision for
students learning programming and computational methods in Python. We have
implemented, used and refined this system since 2009 for growing student
numbers, and summarise the design and experience of using it. The core idea is
to use a unit testing framework: the teacher creates a set of unit tests, and
the student code is tested by running these tests. With our implementation,
students typically submit work for assessment, and receive feedback by email
within a few minutes after submission. The choice of tests and the reporting
back to the student is chosen to optimise the educational value for the students. The system very significantly reduces the staff time required to establish whether a
student's solution is correct, and shifts the emphasis of computing laboratory
student contact time from assessing correctness to providing guidance. The self-paced
nature of the automatic feedback provision supports a student-centred learning
approach. Students can re-submit their work repeatedly and iteratively
improve their solution, and enjoy using the system. We include an evaluation of the system and data from using it in a class of 425 students.
\end{abstract}

\begin{IEEEkeywords}
Automatic assessment tools, automatic feedback provision, programming education, Python, self-assessment technology
\end{IEEEkeywords}

\section{Introduction}

\subsection{Context}

Programming skills are key for software engineering and computer
science but increasingly relevant for computational science outside
computer science as well, for example in engineering, natural and
social science, mathematics and economics. The learning and teaching of
programming is a critical part of a computer science degree and
becoming more and more important in taught and research degrees of
other disciplines.

This paper focuses on an automatic submission, testing and feedback
provision system that has been designed, implemented, used and further
developed at the University of Southampton since 2009 for
undergraduate and postgraduate programming courses. While in this
setting, the primary target group of students were engineers, the same
system could be used to benefit the learning of computer science
students.

\subsection{Effective teaching of programming skills}

One of the underpinning skills for computer science, software
engineering and computational science is programming. A thorough
treatment of the existing literature on teaching introductory
programming was given by Pears \emph{et al.}~\cite{Pears2007}, while a
previous review focused mainly on novice programming and topics
related to novice teaching and learning~\cite{Robins2003}.
Here, we motivate the use of an automatic assessment and feedback system in
the context of teaching introductory programming skills.

Programming is a creative task: given the constraints of the
programming language to be used, it is the choice of the programmer
what data structure to use, what control flow to
implement, what programming paradigm to use, how to name variables and
functions, how to document the code, and how to structure the code
that solves the problem into smaller units (which potentially could be
re-used). Experienced programmers value this freedom and gain
satisfaction from developing a `beautiful' piece of code or finding an
`elegant' solution. For beginners (and teachers) the variety of
`correct' solutions can be a challenge.

Given a particular problem (or student exercise), for example to compute the
solution of an ordinary differential equation, there are a number of criteria
that can be used to assess the computer program that solves the problem:

\begin{enumerate}

\item correctness: does the code produce the correct answer? (For
  numerical problems, this requires some care: for the example of the
  differential equation, we would expect for a well-behaved
  differential equation that the numerical solution converges towards
  the exact solution as the step-width is reduced towards zero.)
  \label{item:aim-correctness}

\item execution time performance: how fast is the solution computed?
  \label{item:aim2}

\item memory consumption: how much RAM is required to compute the
  solution?
  \label{item:aim3}

\item robustness: how robust is the implementation with respect to
  missing/incorrect input values, etc?
  \label{item:aim4}

\item elegance, readability, documentation: how long is the code? Is
  it easy for others to understand? Is it easy to extend? Is it well
  documented, or is the choice of algorithm, data structures and
  naming of objects sufficient to document what it does?
  \label{item:aim-higher-level}

\end{enumerate}

The first aspect -- correctness -- is probably most important: it is better to
have a slow piece of code that produces the correct answer, than to have one
that is very fast but produces a wrong answer. When teaching and providing
feedback, in particular to beginners, one tends to focus on correctness of the
solution. However, the other criteria \ref{item:aim2} to
\ref{item:aim-higher-level} are also important.

We demonstrate in this paper that the assessment of criteria
\ref{item:aim-correctness} to \ref{item:aim4} can be automated in day-to-day teaching of large groups of student.
While the higher-level
aspects such as elegance, readability and documentation of item~\ref{item:aim-higher-level} do require manual inspection of
the code from an experienced programmer, we find that the teaching of the high
level aspects benefits significantly from automatic feedback as all the contact time with
experienced staff can be dedicated to those points, and no time is required to
check the criteria \ref{item:aim-correctness} to \ref{item:aim4}.

\subsection{Automatic feedback provision and assessment}

Over the past two decades interest has been rapidly growing in
utilising new technologies to enhance the learning and feedback
provision processes in higher education. In 1997, Price and Petre
considered the importance of feedback from an instructor to students
learning programming, especially looking into how electronic
assignment handling can contribute to Internet-based teaching of
programming~\cite{Price1997}. Their study compares feedback given
manually by several instructors to cohorts of conventional and
Internet learning students, only a small fraction of which involved
running the students' submissions. For the functional programming
language Scheme, Saikkonen \emph{et al.} described a system that
assesses programming exercises with the possibility to analyse
individual procedures and metrics such as run
time~\cite{Saikkonen2001}. A feedback system called ``submit'' for
code in Java was introduced in 2003, which worked by allowing users to
upload code, which would be compiled and (if the compilation was
successful) run, with the output displayed for comparison with model
output provided by the lecturer; the lecturer would manually grade the
work later, and the system would also display this
information~\cite{Venables2003}. Recognising the popularity of
test-driven development and adopting that approach in programming
courses, Stephen Edwards implemented a system, web-CAT, that would
assess both the tests and the code written by
students~\cite{Edwards2003}. Shortly thereafter, another group
produced a tool for automatically assessing the style of C++
programs~\cite{Ala-Mutka2004}, which students were encouraged to use,
and which was also used be instructors when manually assessing
assignments; it was found that the students started to follow many
important style guidelines once the tool was made available.

By 2005 there was sufficient interest in the field of automatic
assessment systems that multiple reviews were
published~\cite{Douce2005,Ala-Mutka2005}, highlighting
the emergence of evidence that automatic assessment
can lead to increased student
performance~\cite{Woit2003,English2002}. Another benefit realised with
automatic assessment systems is greater ease in detecting plagiarism,
tools for the purpose having been included in several of the systems
surveyed. Also reported on that year was
CourseMarker~\cite{Higgins2005}, which can mark C++ and Java programs,
and uses a Java client program to provide a graphical user interface
to students.

A more recent review of automatic assessment
systems~\cite{Ihantola2010} which highlighted newer development
recommended that future systems devote more attention to security, and
future literature describe more completely how the systems work.
A work from MIT CSAIL and Microsoft
introduces a model in which the system -- provided with a
reference implementation of a solution, and an error model consisting
of potential correction to errors that students may make --
automatically derives minimal corrections to students' incorrect
solutions~\cite{Singh2013}.
Another relatively recent development is the
adoption of distributed, web-based training and assessment
systems~\cite{Verdu2012}, as well as the increasingly-popular
``massive open online courses'' or MOOCs~\cite{Masters2011}.
A current innovation in the field is the nbgrader
project~\cite{nbgrader}, an open-source project that is designed for
generating and grading assignments in IPython notebooks~\cite{Perez2007}.

\subsection{Outline}

In this work, we describe motivation, design, implementation and
effectiveness of an automatic feedback system for Python programming
exercises used in undergraduate teaching for engineers. We aim to
address the shortcomings of the current literature as outlined in the
review \cite{Ihantola2010} by detailing our implementation and
security model, as well as providing sample testing scripts, inputs
and outputs, and usage data from the deployed system. We combine the
provision of the technical software engineering details of the testing
and feedback system, with motivation and explanation of its use in a
educational setting, and data on student reception based on 6 years of
experience of employing the system in multiple courses and countries.

In Sec.~\ref{sec:trad-deliv-progr}, we provide some historic context
of how programming was taught prior the introduction of the automatic
testing system described here. Sec.~\ref{sec:new-method-automatic}
introduces the new method of feedback provision, initially from the
student's perspective -- who are the users from a software engineering
point of view -- then providing more detail on design and
implementation. Based on our use of the system over multiple years, we
have composed results, statistics and a discussion of the system in
Sec.~\ref{sec:results}, before we close with a summary in
Sec.~\ref{sec:summary}.

\section{Traditional delivery of programming education}
\label{sec:trad-deliv-progr}

In this section, we describe the learning and teaching methods used in
the Engineering degree programmes at the University of Southampton
before the automatic feedback system was introduced.

\subsection{Programming languages used}

We taught languages such as C and MATLAB to students in Engineering as
their first programming languages until 2004, when we introduced
Python~\cite{Fangohr2005} into the curriculum. Over time,
we have moved to teaching Python as a versatile
language~\cite{Griffiths2009,Bogdanchikov2013} that is relatively easy
to learn~\cite{Fangohr2004} and useful in wide variety of
applications~\cite{PythonApps, Fangohr2006}. We teach C for advanced students
in later years as a compiled and fast language.

\subsection{Lectures}\label{sec:lectures}

Lectures that introduce a programming language to beginners are
typically scheduled over a duration of 12 weeks, with two 45 minute
lectures per week. This is combined with a scheduled computing
laboratory (90 minutes) every week (Sec.~\ref{sec:comp-labor}), and an
additional and optional weekly ``help session``
(Sec.~\ref{sec:help-session})

The lectures introduce new material, demonstrate what one can do with
new commands, and how to use programming elements or numerical
methods. In nearly all lectures, new commands and features are used
and demonstrated by the lecturer in live-coding of small programs;
often with involvement of the students. The lectures are thus a
mixture of traditional lectures and a tutorial-like component where
the new material is applied to solve a problem, and -- while only the
lecturer has a keyboard which drives a computer with display output
connected to a data projector -- all students contribute, or are at
least engaged, in the process of writing a piece of code.

\subsection{Computing laboratories}\label{sec:comp-labor}

However, for the majority of students the actual learning takes
place when they carry out programming exercises themselves.

To facilitate this, computer laboratory sessions (90 minutes every week) are
arranged in which each student has one computer, and works at their own pace
through a number of exercises. Teaching staff are available during
the session, and we have found
that about 1 (teaching assistant) demonstrator per 10 students is required for this set up.

The lecturer and demonstrators (either academics or
postgraduate students) fulfil three roles in these laboratory
sessions:
\begin{enumerate}
\item[(i)] to provide help and advice when students have
difficulties or queries while carrying out the self-paced exercises,
\item[(ii)] to establish whether a student's work is correct (i.e. does the
student's computer program do what it is meant to do), and
\item[(iii)] to
provide feedback to the student (in particular: what they should
change for future programs they write).
\end{enumerate}

Typically, prior to introducing the automatic testing system in 2009, the
teaching assistants were spending 90\% of their time on activity (ii), i.e.
checking students' code for correctness, and the remaining 10\% of
time can be used on (i) and (iii), while the educational value is
overwhelmingly in (i) and (iii).

In practical terms, the assessment and feedback provision was done in pairs
consisting of one demonstrator and one student looking through the student's
files on the student's computer at some point during the subsequent computing
laboratory session. The feedback and assessment was thus delivered one week
after the students had completed the work.

\subsection{Help session}\label{sec:help-session}

In the weekly voluntary help session, computers and teaching staff
are available for students if they need support exceeding the normal
provision, would like to discuss their solutions in more depth, or
seek inspiration and tasks to study topics well beyond the expected
material.

\section{New Method of automatic feedback provision}

\label{sec:new-method-automatic}

\subsection{Overview} In 2009, we introduced an \emph{automatic feedback
provision system} that checks each student's code for correctness and provides
feedback to the student within a couple of minutes of having completed the
work. This takes a huge load off the demonstrators who consequently can spend
most of their time helping students to do the exercises (item (i) in
Sec.~\ref{sec:comp-labor}) and providing additional feedback on completed and
assessed solutions (item (iii) in Sec.~\ref{sec:comp-labor}). Due to the
introduction of the system the learning process can be supported considerably
more effectively, and we could  reduce the number of demonstrators from 1 per
10 students as we had pre-2009, to 1 demonstrator per 20 students, and still
improve the learning experience and depth of material covered. There was no
change to the scheduled learning activities, i.e. the weekly lectures
(Sec.~\ref{sec:lectures}), computing laboratory sessions (Sec.~\ref{sec:comp-labor}),
and help sessions (Sec.~\ref{sec:help-session}) remain.

In Sec.~\ref{sec:stud-persp}
``Student's perspective'' we show a typical example of a very simple
exercise, along with correct and incorrect solutions, and the feedback
that those solutions give rise to.
Later sections detail the system design and work flow
(Sec.~\ref{sec:design-and-implementation}) and in particular the
implementation of the student code
testing (Sec.~\ref{sec:impl-test}), with reference to this example exercise.

\subsection{Student's perspective\label{sec:stud-persp}}
Once a student completes a programming exercise in the computing
laboratory session, they send an email to a dedicated email account
that has been created for the teaching course, and attach the file
containing the code they have written. The subject line is used by the
student to identify the exercise; for example ``Lab 4`` would identify
the 4$^\mathrm{th}$ practical session. The system receives the
student's email, and the next thing that the student sees is an
automatically generated email confirmation of the submission (or,
should the submission not be valid, an error message is emailed
instead, explaining why the submission was invalid. Invalid submission
can occur for example if emails are sent from email accounts that are
not authorised to submit code). At this stage, the student's code is
enqueued for testing, and after a short interval, the student receives
another email containing their assessment results and feedback by email.  Where
problems are detected, this email also includes details of what the
problems were. Typically, the student will receive feedback in their
inbox within two to three minutes of sending their email.

We shall use the following example exercise, which is typical of
one that we might use in an introductory Python laboratory, as
the basis for our case study:\\
\fcolorbox{black}{lightgrey}{%
  \begin{minipage}{\columnwidth} Please define the following functions in the
    file training1.py and make sure they behave as expected. You also
    should document them suitably. \begin{enumerate}
    \item A function \texttt{distance(a, b)} that returns the distance
      between numbers \texttt{a} and \texttt{b}.
    \item A function \texttt{geometric\_mean(a, b)} that returns the
      geometric mean of two numbers, \emph{i.e.} the edge length that
      a square would have so that its area equals that of a rectangle
      with sides \texttt{a} and \texttt{b}.
    \item A function \texttt{pyramid\_volume(A, h)} that computes and
      returns the volume of a pyramid with base area \texttt{A} and
      height \texttt{h}.
\end{enumerate}
\end{minipage}}

\vspace{0.5cm} We show a correct solution to question 3 of this
example exercise in~\lstref{lst:corr}. If a student who is enrolled on
the appropriate course submits this, along with correct responses to
the other questions, by email to the system, they will receive
feedback as shown in~\lstref{lst:corr-res}.

\begin{lstlisting}[language=python,float=htb!,label={lst:corr},caption=A correct solution to question 3 of the example exercise]
def pyramid_volume(A, h):
    """Calculate and return the volume of a pyramid
    with base area A and height h.
    """
    return (1./3.) * A * h
\end{lstlisting}

\begin{lstlisting}[float=hbt!,label={lst:corr-res},caption={email response to correct submission, additional line wrapping due to column width},language=]
Dear Neil O'Brien,

Testing of your submitted code has been completed:

Overview
========

test_distance       : passed -> 100% ; with weight 1
test_geometric_mean : passed -> 100% ; with weight 1
test_pyramid_volume : passed -> 100% ; with weight 1

Total mark for this assignment: 3 / 3 = 100%.

(Points computed as 1 + 1 + 1 = 3)

-----------------------------------------------

This message has been generated  automatically. Should
you feel that you observe a malfunction of the system,
or if you wish to speak to a human, please contact the
course team (course-help@uni.email.address).
\end{lstlisting}

If the student submits an incorrect solution, for example with a
mistake in question~3 as shown in~\lstref{lst:wrong}, they will
instead receive the feedback shown in~\lstref{lst:wrong-res}. Of
course the students must learn to interpret this style of feedback in
order to gain the maximum benefit, but this is in itself a useful
skill, as we discuss more fully
in~\secref{sec:quality-of-feedback-provision}, and comments from the
testing code assist the students, as discussed
in~\secref{sec:results-fb-provision}. The submission
in~\lstref{lst:wrong} is incorrect because integer division is used
rather than the required floating-point division. These exercises
are based on Python 2, where the ``/'' operator represents
integer division if both operands are of integer type, as is common in
many programming languages (in Python 3, the ``/'' operator represents
floating point division even if both operands are of type integer).

\begin{lstlisting}[language=python,float=hbt!,label={lst:wrong},caption={An incorrect solution to question 3 of the example exercise, using integer division}]
def pyramid_volume(A, h):
    """Calculate and return the volume of a pyramid
    with base area A and height h.
    """
    return (A * h) / 3
\end{lstlisting}

\begin{lstlisting}[float=hbt!,breaklines,label={lst:wrong-res},caption={email response to incorrect solution},language=]
Dear Neil O'Brien,

Testing of your submitted code has been completed:

Overview
========

test_distance       : passed -> 100% ; with weight 1
test_geometric_mean : passed -> 100% ; with weight 1
test_pyramid_volume : failed ->   0% ; with weight 1

Total mark for this assignment: 2 / 3 = 67%.

(Points computed as 1 + 1 + 0 = 2)

Test failure report
====================

test_pyramid_volume
-------------------
def test_pyramid_volume():

    # if height h is zero, expect volume zero
    assert s.pyramid_volume(1.0, 0.0) == 0.

    # tolerance for floating point answers
    eps = 1e-14

    # if we have base area A=1, height h=1,
    # we expect a volume of 1/3.:
    assert abs(s.pyramid_volume(1., 1.) - 1./3.) < eps

    # another example
    h = 2.
    A = 4.
    assert abs(s.pyramid_volume(A, h) -
               correct_pyramid_volume(A, h)) < eps

    # does this also work if arguments are integers?
>   assert abs(s.pyramid_volume(1, 1) - 1. / 3.) < eps
E   assert 0.3333333333333333 < 1e-14
E    + where 0.3333333333333333 = abs((0 - (1.0/3.0)))
E    +   where 0 = <function pyramid_volume at
                    0x7f0ce1af4e60>(1, 1)
E    +     where <function pyramid_volume at
                  0x7f0ce1af4e60> = s.pyramid_volume
\end{lstlisting}

Within the testing feedback in \lstref{lst:wrong-res}, the student code is
visible in the name space \verb+s+, i.e. the function \verb+s.pyramid_volume+
is the function defined in \lstref{lst:wrong}. The function
\verb+correct_pyramid_volume+ is visible to the test system but students
cannot see the implementation in the feedback their receive -- this allows us to define tests that
compute complicated values for comparison with those computed by the student's submission, without
revealing the implementation of the reference computation to the students.

\subsection{Design and Implementation}\label{sec:design-and-implementation}

The design is based on three different processes that are started periodically (every minute) and communicate via file system based task queues with each other:
\begin{enumerate}
\item A \emph{incoming queue} of incoming student submissions, initial
  validation and extraction of files and required tests to run (see
  high level flow chart in Fig.~\ref{fig:flowdiagram-incoming})
\item A \emph{queue of outgoing messages} that need to be delivered to
  the users and administrators which -- in our email based user
  interface -- decouples the actual testing queue from availability of
  the email servers (flow chart in
  Fig.~\ref{fig:flowdiagram-outgoing}).
\item A \emph{queue of tests} to be run, where the actual testing of
  the code takes place in a restricted environment (flow chart in
  Fig.~\ref{fig:flowdiagram-testing})
\end{enumerate}

We describe how these work together in more detail in the following sections.

The system is implemented in Python, and primarily tests Python code
(in~\secref{sec:oth-lang} we discuss generalisation of the system to
test code in other languages).

\subsubsection{Email receipt and incoming queue process\label{sec:receipt-decoding}}

Each course that uses the automatic feedback provision system has a
dedicated email account set up to receive submissions. At the
University of Southampton, for a course with code ABC,
the email address would be \url{ABC@uni.email.address}. As the
subject line, the student has to use a predefined string (such as
\texttt{lab 1}), which is specified in the assignment instructions, so
the testing system can identify which submission this is. The identity
of the student is known through the email address of the sender.

\begin{figure*}
  \centering
  \begin{subfigure}[b]{1\textwidth}
      \includegraphics[width=\textwidth]{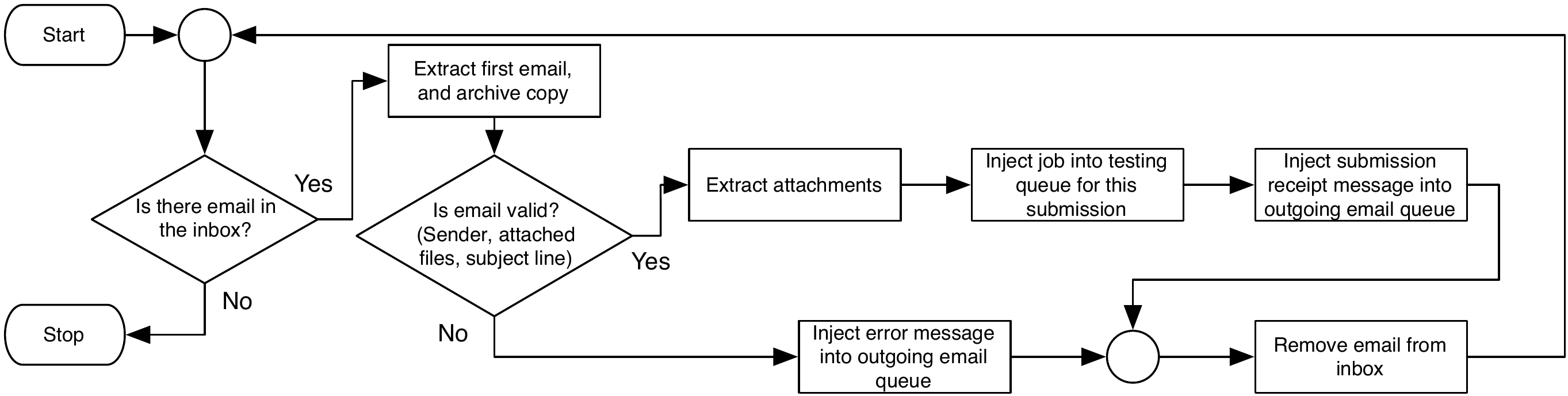}
    \caption{Incoming queue process\label{fig:flowdiagram-incoming}}
  \end{subfigure}
  \medskip

  \begin{subfigure}[b]{1\textwidth}
    \includegraphics[width=\textwidth]{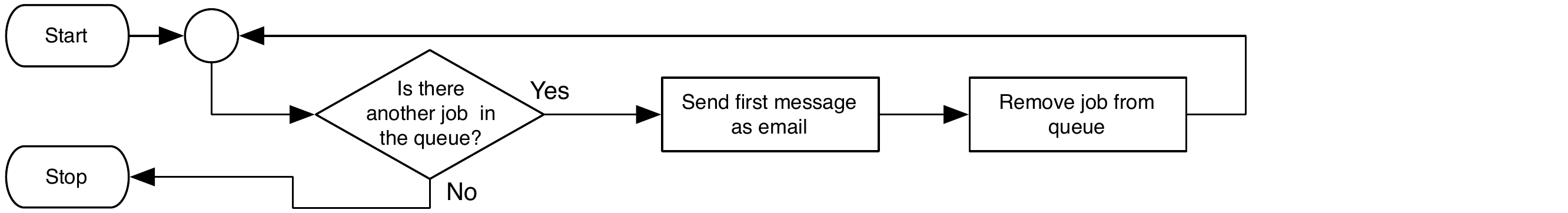}
    \caption{Outgoing email queue process\label{fig:flowdiagram-outgoing}}
  \end{subfigure}
  \medskip

  \begin{subfigure}[b]{1\textwidth}
    \includegraphics[width=\textwidth]{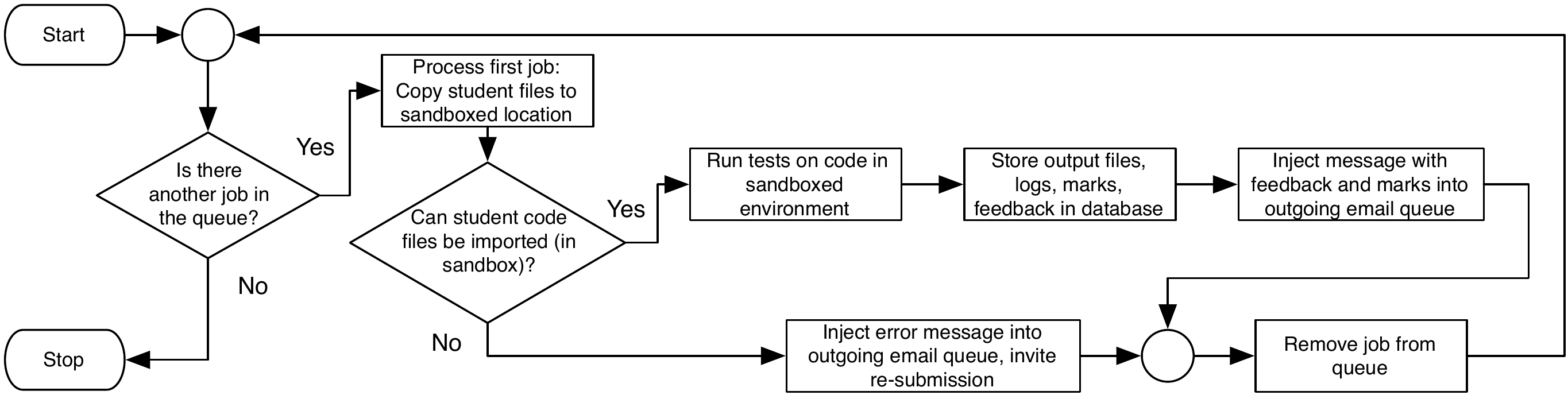}
    \caption{Testing queue process\label{fig:flowdiagram-testing}}
  \end{subfigure}
  \caption{Flow charts illustrating the work flow in each
    process. Processes are triggered every minute via a cronjob entry,
    and don't start until their previous instance has
    completed. \label{fig:flowdiagrams}}

\end{figure*}

The testing system accesses the email inbox every minute, and
downloads all incoming mails from it using standard tools such as
\texttt{fetchmail}, or \texttt{getmail} combined with \texttt{cron}.  These incoming mails are then
processed sequentially as summarised in the flow chart in
Fig.~\ref{fig:flowdiagram-incoming}:
\begin{enumerate}
\item The email is copied, for backup purposes, to an archive of all
  incoming mail for the given course and year.
\item The email is
  checked for validity in the following respects:
  \begin{enumerate}
  \item the student must be known on this course (this is checked
    using a list of students enrolled on the course, provided by the
    student administration office); submissions from students who are
    not enrolled are logged for review by an administrator in case the
    student list was not correct or a student has transferred between
    courses;
  \item the subject line of the email must relate to a valid exercise
    for the course;
  \item all required files must be attached to the email, and these
    must be named as per the instructions for the exercise.
  \end{enumerate}
\item If the email is invalid (i.e., one or more of the above criteria
  are not met), an error report is created and enqueued in the outgoing email queue for delivery. The email explains why the
  submission is not valid, inviting the student to correct the
  problems and re-submit their work.
\item For a valid submission, the attachments of the incoming email containing the student's
  code are saved and
\item an item is placed into the testing queue,
  including the exercise that is to be tested, the student's user
  name, and names and paths of the files that were submitted.
  \item For a valid submission, an email to the student is enqueued in the outgoing message queue that confirms receipt of the submission; the student can use this to evidence their submission and submission time, and it re-assures the students that all required files were present, and that the submission has entered the system.
  \item For both valid and invalid submissions, the email is removed from the incoming queue.
\end{enumerate}

\subsubsection{Outgoing messages\label{sec:outgoing-messages}}

The implementation of sending error messages and feedback reports to the students, and
any other messages to administrators, is realised through a separate queue and process for
outgoing messages (see Fig.~\ref{fig:flowdiagram-outgoing} and discussion of this design in Sec.~\ref{sec:resilience}). This process is also used for weekly emails informing students about the overall progress (Sec.~\ref{sec:statistical-reporting-to-lecturers-and-students}).

We note in passing that all automatically generated messages invite the
student to contact the course leader, other teaching staff or the
administrator of the feedback provision system should they not understand the
email or feel that the system has malfunctioned; help can be sought by email or in person during
the timetabled teaching activities.

\subsection{Design and implementation of student code
  testing\label{sec:impl-test}}

The testing queue shown in Fig.~\ref{fig:flowdiagram-testing} processes
submissions that have been enqueued by the incoming mail processing script. The task is
to execute a number of predefined tests against the student code in a secure
environment, using unit testing tools
to establish correctness of the student submission. As we use Python
for these courses in computation for science and engineering, we can
plug into the testing capabilities that come
with Python, and those that are provided by third party tools, such as
nose~\cite{nose} and pytest~\cite{pytest}. We have chosen the
\verb+py.test+ tool because we have more experience with this system.

Here, we provide a brief overview of the testing process which is invoked
every minute (unless an instance started earlier has not completed execution
yet), with Sections.~\ref{sec:iterative-tests} to
\ref{sec:last-subsubsection-in-testing-implementation} providing more
details on the requirements and
chosen design and implementation.

For each testing job found in the queue, the following steps are carried out (see the flow diagram in Fig.~\ref{fig:flowdiagram-testing}):
\begin{enumerate}
\item The student files to be tested are copied to a sand-boxed location on the file system with limited access permissions (Sec.~\ref{sec:security}).
\item A dedicated local user with minimal privileges tries to import the code in a Python process to check for correct syntax.
\item If the import fails due to syntax errors an error message is prepared for the user and injected into the outgoing message queue. (See also Sec.~\ref{sec:Submissions-including-syntax-errors} for a discussion.) The job is removed from the testing queue and the process moves to the next item in the queue.
\item If the import succeeds, the tests are run on the submitted code in the restricted environment (Sec.~\ref{sec:iterative-tests} to \ref{sec:clean-code-and-PEP8}).
\item Output files (that the student code may produce) and testing logs are archived, marks extracted and all data are stored in a database which may be used by the lecturer to discover the marks for each student, for each question and assignment.
\item A feedback message for the student is prepared and injected into the outgoing message queue containing the test results (Sec.~\ref{sec:results-fb-provision}). This provides the student with a score for each question in the assignment, and where mistakes were found, provides details of the particular incorrect behaviour that was discovered.
\lstref{lst:wrong-res} shows an example of such feedback.
\item The test job is removed from the queue.
\end{enumerate}

We discussion additional weekly feedback to students in Sec.~\ref{sec:statistical-reporting-to-lecturers-and-students} and the system's dependability in Sec.~\ref{sec:resilience}.

\subsubsection{Security measures\label{sec:security}}

By the nature of the testing system, it contains student data (names,
email addresses, and submissions), and it is incumbent upon the
developers and administrators to take all reasonable measures to
safeguard these data against unauthorised disclosure or modification.
We also require the system to maintain a high availability and
reliability.  The risks that we need to guard against can largely be
divided into two categories: (i) genuine mistakes made by students in
their code, and (ii) attempts by students -- or others who have
somehow gained access to a student's email account -- to intentionally
access or change their own or other students' work, assigned marks, or
other parts of the testing system.

Experience shows that some of the most common genuine mistakes made by
students include cases such as unterminated loops, which would execute
indefinitely.  Due to the serialisation of the tests in our system,
this problem, if left unchecked, would stop the system processing any
further submissions until an administrator corrected it.  However, we
have applied a POSIX resource limit~\cite{POSIX,APUE3} on CPU time to
ensure that student work consuming more than a reasonable and fixed
limit is terminated by the system.  We catch any such terminations, and in this case we have adopted a policy of
informing the student by email, and giving them the opportunity to
re-submit an amended version of their work.  We apply similar resource
limits on both disk space consumption and virtual memory size, in
order that loops which would output large amounts of data to
\verb+stdout+, \verb+stderr+, or a file on disk, or which interminably
append to a list or array resulting in its consumption of unreasonable
amounts of memory, are also prevented from causing an undue impact on
the testing machine's resources.

We address the potential that submitted code could attempt to
maliciously access data about another student (or parts of the system)
with a multi-faceted approach:
\begin{enumerate}
\item We execute the tests on the student code under a separate local user
  account on the server that performs the tests.  This account has
  minimal permissions on the file system.
\item We create a separate directory for each submission that we test,
  and run the tests within this directory.
\item The result of the two previous points, assuming that all
  relevant file system permissions are configured correctly, means
  that no student submission may read or modify any other student's
  submissions or marks, nor can it read the code comprising the
  testing system.
\item The environment variables available to processes running as the
  test user are limited to a small set of pre-defined variables, so
  that no sensitive data will be disclosed through that mechanism.
\item We do not provide the students information about the file system
  layout, local account names, etc. on the host that runs the tests,
  to reduce the chance that students know of the locations of
  sensitive data on the file system.
\end{enumerate}

\subsubsection{Iterative testing of student code\label{sec:iterative-tests}}

We have split the exercises on our courses into questions, and
arranged to test each question separately.  Within a question, the
testing process stops if any of the test criteria are not satisfied.
This approach was picked to encourage an iterative process whereby
students are guided to focus on one mistake at a time, correct it, and
get further feedback, which improves the learning experience.  This
approach is similar to that taken by Tillmann \emph{et
  al.}~\cite{Tillmann2013}, where the iterative process of supplying
code that works towards the behaviour of a model solution for a given
exercise is so close to gaming that it ``is viewed by users as a game,
with a byproduct of learning''. Our process resembles test-driven
development strategies and familiarises the students with test-driven
development \cite{TDD} in a practical way.

\subsubsection{Defining the tests\label{sec:tests}}

There are an indefinite number of both correct and incorrect ways to
answer an exercise, and to test correctness using a regression testing
framework requires some skill and experience in constructing a
suitably rigorous test case for the exercise. We build on our
experience before and after the introduction of the testing system,
ongoing feedback from interacting with the students and reviewing
their submissions to design the best possible unit testing for the
learning experience. This includes testing for correctness but also
structuring tests in a didactically meaningful order.  Comments added
in the testing code will be visible to the students when a test fails,
and can be used to provide guidance to the learners as to what is
tested for, and what the cause of any failure may be (if desired).

Considering question 3) in the example exercise we introduced
in~\secref{sec:stud-persp}, the tests that we carry out on the
student's function include the following:
\begin{enumerate}
\item Volume must be 0 when \verb+h+ is 0.
\item Volume must be 0 when \verb+A+ is 0.
\item If we have \verb+A = 1+ and \verb+h = 3+, volume must be 1.
\item If we have \verb+A = 3+ and \verb+h = 1+, volume must be 1.
\item If we have \verb+A = 1.0+ (as a float) and \verb+h = 1.0+ (as a
  float), volume must be $\frac{1}{3}$.
\item If we test another combination of values of floating-point
  numbers \verb+A+ and \verb+h+ then the returned volume must be
  \verb+A * h / 3.0+.
\item If we have \verb+A = 1+ (as an integer) and \verb+h = 1+ (as an
  integer), volume must be $\frac{1}{3}$.
\item The function must have a documentation string; this must contain
  several words, one of which is ``return''.
\end{enumerate}

In this very simple example, we set up the first group of criteria
(1--6) to determine that the student has implemented the correct
formula to solve the problem at hand.  Criterion 7 tests for the
common mistake of using integer division where floating-point division
is required.  The final criteria concern coding style.  In this
example, it is a strict requirement that the code is documented to at
least some minimal standard, and the student will gain no marks for a
question that is answered without a suitable documentation string.

\begin{lstlisting}[language=python,float=htb!,label={lst:testdef},caption={testing code for example question}]
def test_pyramid_volume():

    # if height h is zero, expect volume zero
    assert s.pyramid_volume(1.0, 0.0) == 0.

    # if base A is zero, expect volume zero
    assert s.pyramid_volume(0.0, 1.0) == 0.

    # if base has area A=1, and the height is h=3,
    # we expect a volume of 1:
    assert s.pyramid_volume(1.0, 3.0) == 1.

    # if base has area A=3, and the height is h=1,
    # we expect a volume of 1:
    assert s.pyramid_volume(3.0, 1.0) == 1.

    #acceptable tolerance for floating point answers
    eps = 1e-14

    # if base has area A=1, and the height is h=1,
    # we expect a volume of 1/3.:
    assert abs(s.pyramid_volume(1., 1.) - 1./3.) < eps

    # another example
    h = 2.
    A = 4.
    assert abs(s.pyramid_volume(A, h) -
               correct_pyramid_volume(A, h)) < eps

    # does this also work if arguments are integers?
    eps = 1e-14
    assert abs(s.pyramid_volume(1, 1) - 1./3.) < eps

    # is the function documented well
    docstring_test(s.pyramid_volume)
\end{lstlisting}

Our implementation of the tests described above is given
in~\lstref{lst:testdef}.  In implementing these criteria, we avoid
testing for exact equality of floating point numbers at any point in
the testing process.  Instead we define some tolerance
(e.g. \verb+eps = 1e-14+), and require that the magnitude of the
difference between the result of the student's code and the required
answer be below this tolerance.  This avoids failing student
submissions which have e.g. performed accumulation operations in a
different order and concomitantly suffered differing floating-point
round-off effects. As exercises become more complex and related to
numerical methods, a different tolerance may have to be chosen.

We order the criteria so those that are most likely to pass are tested
earlier, and we have chosen to stop the testing process at the first
error encountered.  This encourages students to address and correct
one error at a time in an iterative process, if required, which is
possible thanks to the short timescale between their submitting work
and receiving feedback (see Sec.~\ref{sec:iterative-tests}).  The
implementation of the tests for \verb+py.test+ is based on
\verb+assert+ statements, which are \verb+True+ when the student's
code passes the relevant test, and \verb+False+ otherwise.  The final
criteria, that the documentation string must exist and pass certain
tests, is handled by asserting that a custom function that we provide
to check the documentation string returns \verb+True+.  Of course, the
tests must be developed carefully to suit the exercise they apply to,
and to exercise any likely weaknesses in the students' answers, such
as the chance that integer division would be used in the
implementation of the formula for the volume of a pyramid discussed
above.

\subsubsection{Clean code and PEP~8}\label{sec:clean-code-and-PEP8}

In addition to the hard syntactic requirements of a programming
language, there are often recommendations how to style and lay-out
source code. We find that it is very efficient to introduce this to
students from the very beginning of their programming learning
journey.

For Python, the so-called ``PEP~8 Style Guide'' for Python
Code~\cite{PEP8} is useful guidance, and electronic tools are
available to check that code follows these voluntary recommendations
for clean code. PEP~8 has recommendations for the number of spaces
around operators, before and after commas, the number of empty lines
between functions, class definitions, etc.

We use the \verb+pep8+ utility \cite{pep8tool} to assess the
conformance of the student's entire submission file (which will
usually consist of answers to several questions like the above) with
the PEP~8 Style Guide.  Our system counts the number of errors that
are found, $N_{\mathrm{err}}$, and penalises the student's total score
according to a policy (e.g. we may choose a policy of multiplying the
raw mark that could be obtained for full PEP~8 conformance by
$2^{-N_{\mathrm{err}}}$, or of implementing any other desired mark
adjustment as a function of that value).

\subsubsection{Results and feedback provision to student\label{sec:results-fb-provision}}

The results of the testing process are written to machine-readable
files by \verb+py.test+.  For each tested submission, the report is
parsed by our system, with one of a number of results being possible:
the student code may have run completely in which case, we
have a pass result or a fail result for each of the defined tests.
Otherwise the student code may have terminated with an error which is
most likely due to a resource limit being exceeded causing
the operating system to abort the process, as discussed
in~\secref{sec:security}.

The number of
questions that were answered correctly (i.e. have no failed assertions
in the associated tests) is counted and stored in a database.  If
there were incorrect answers, we extract a backtrace from the
\verb+py.test+ output which we incorporate into the email that is sent
into the student.  The general format of the results email is to give
a per-question mark, with a total mark for the submission, and then to
detail any errors that were encountered.
In the calculation of the mark for the assessment, questions can be given
different weights to reflect greater importance or challenges of
particular questions. For the example shown in \lstref{lst:corr-res}
all questions have the same weight of~1.

We described and illustrated a typical question, which might form part
of an assignment, in~\secref{sec:stud-persp}.  As shown in~\lstref{lst:wrong-res}, when an error is encountered, the
results that are sent to the student include the portions of the
testing code for the question in which the error was found that have
passed successfully, and then indicate with the \verb+>+ character the
line whose assertion failed (in this case the 7th-last line shown).
This is followed by a backtrace which illustrates that, in this case,
the submitted \verb+pyramid_volume+ function returned \verb+0+ when it
was expected to return an answer of $\frac{1}{3}\pm1\times10^{-14}$.
The report also includes several comments, which are
introduced in the testing code (shown in~\lstref{lst:testdef}), and
assist students in working out what was being tested when the error
was found.  Here, the comment ``does this also work if arguments are
integers?'' shows the student that we are about to test their work
with integer parameters; that should prompt them to check for integer
division operations.  If they do not succeed in doing this, they will
be able to show their feedback to a demonstrator or academic, who can
can use the feedback to locate the error in the student's code
swiftly, then help the student find the problem, and discuss ways
to improve the code.

\subsubsection{Statistical reporting to lecturers and routine
  performance feedback to student}\label{sec:statistical-reporting-to-lecturers-and-students}
The system records all pertinent data about each submission including
the user who made the submission, as well as the date and time of the
submission, and the mark awarded.  We use this data to further engage
students with the learning process, by sending out a weekly email
summary of their performance to date, as shown in~\lstref{lst:reg-fb}.
This includes a line for each exercise whose deadline has passed,
which reminds the student of their mark and whether their submission
was on time or not.  For a student who has submitted no work, a
different reminder is sent out, requesting they submit work, and
giving contact details of the course leader, asking them to make
contact if they are experiencing problems. Messages are sent via the outgoing queue (Sec.~\ref{sec:outgoing-messages}).

\begin{lstlisting}[float=hbt!,breaklines,label={lst:reg-fb},caption={Typical routine feedback email},language=]
Dear Neil O'Brien,

Please find below your summary of submissions and
preliminary marks for the weekly laboratory sessions
for course ABC, as of Fri Jan 30 17:06:44 2015.

lab  2 :  25%   Details:  1.00 /  4.00,
                          submitted before deadline
lab  3 :  31%   Details:  1.25 /  4.00,
                          submitted before deadline
lab  4 :   0%   Details:  4.00 /  4.00, but submission
                          at 2014-11-14 20:39:02 was
                          late by 4:39:02.
lab  5 :  80%   Details:  4.00 /  5.00,
                          submitted before deadline
lab  6 :  77%   Details:  3.06 /  4.00,
                          submitted before deadline
lab  7 :  75%   Details:  3.00 /  4.00,
                          submitted before deadline

The average mark over the listed labs is 48%.

With kind regards,

The teaching team (course-help@uni.email.address)
\end{lstlisting}

We also monitor missing submissions in the first couple of weeks very
carefully and contact students individually who appear not to have
submitted any work. Occasionally, they are registered on the wrong
course, but similarly some students just need a little bit of extra
help with their first ever programming exercises and by expecting the
first submitted work at the end of the first or second teaching week,
we can intervene early in the semester and help those students get
started with the exercises and follow the remainder of the course.

After the deadline for each set of exercises, the course lecturers will
generally flick through the code that students have submitted (or at
least 10 to 20 randomly chosen submissions if the number of students
is large). This helps the teacher in identifying typical patterns and
mistakes in the students' solutions, which can be discussed, analysed
and improved effectively in the next lecture: once all student
specific details are removed from the code (such as name, login and
email address), submitted (and anonymised) code can be shown in the
next lecture.  We find that students clearly enjoy this kind of
discussion and code review jointly carried out by students and
lecturer in the lecture theatre, in particular where there is the
possibility that their anonymised code is being shown (although only
they would know).

The data for the performance of the whole class is made available to
the lecturer through private web pages which allow quick
navigation to each student and all their submissions, files and
results. Key data is also made available as a spreadsheet, and a
number of graphs showing the submission activity (some are shown in
Figs.~\ref{fig:l2-t2-histo}, \ref{fig:tr-subm-v-time} and
\ref{fig:labs-subm-v-time} and discussed in section~\ref{sec:results}).

\subsubsection{Dependability and resilience}
\label{sec:resilience}
\label{sec:last-subsubsection-in-testing-implementation}
The submission system is a critical piece of infrastructure for the
delivery of those courses that have adopted it as their marking
and feedback system; this means that its reliability and availability must be
maximised.  We have taken several measures to reduce the risk of
downtime and service outages, and also to reduce the risk of data loss
to a low level.

The machine on which the system is installed is a virtual machine
which is hosted on centrally managed University infrastructure.  This
promises good physical security for the host machines, and
high-availability features of the hypervisor, such as live
migration~\cite{Mashtizadeh2011}, improve resilience against possible
individual hardware failures.  To combat the possibility of the data
(especially the student submissions) being lost we have instituted a
multi-tier backup system, which backs up the system's data to multiple
physical locations and to multiple destination storage media, so that
the probability of losing data should be very small.

The remaining potential single point of failure is the University's email
system, which is required for any student to be able to submit work or
receive feedback.  In the case that the email system were to fail close to
a deadline, we would have the choice of extending the deadline to
allow submissions after the service was restored, and/or manual
intervention to update marks where students could demonstrate that
their submission was ready on time, depending on the lecturer's chosen
policy.

The internal architecture of the testing system was designed to be as
resilient as possible, and to limit the potential impact of any faults. 
A key approach to this goal is the use of
various (file system based) queues (Fig.~\ref{fig:flowdiagrams}) that
decouple the different stages of submission handling and testing so
that e.g. a failure of the system's ability to deliver emails would
not impede testing submissions already received.  Emails are received
into a local mailbox and are processed one item at a time so this is
the first effective queue; receipt of emails can continue even if
the testing process has halted.  Valid submissions from processed emails are
then entered into a queue for testing, the entries of which are
processed sequentially.  The receipt and testing processes generate
emails, which are placed into an outgoing mail queue and are sent
regularly, the queue items being removed only after successful
transmission.  This way, if the outgoing email service is unavailable,
mails will accumulate in the queue and be sent \emph{en masse} when
the service is restored.

Another key design decision was that each individual part of the
receipt and testing process is carried out sequentially for each
submission and is protected by lock files.  Prior to processing received messages, the systems checks for existing locks; if these exist,
the processing doesn't start, and the event is logged (receipt of other emails continues as the receiving and processing are separate processes).  If no locks
are found, a lock is created, which is removed upon successful
processing; any unexpected termination of the processing code will
result in a lock file being left behind, so that we can investigate
what went wrong and make any required corrections before restarting
the system.  The testing process itself is likewise protected by
locking. A separate watch dog process alerts the administrator if lock files have stayed in place for more than a few minutes -- typically each process completes within a minute.

In practice we have developed the system to the point that
we have not had an unexpected failure require us to manually clean up
and unlock in two years of production use, but should an unexpected
bug be found, this design ensures that at most one submission will be
affected (a copy will have been made before any processing was carried
out, so even in this case there would be no loss of student data).

\section{Results}
\label{sec:results}

\subsection{Testing system deployment}

The automatic testing system was first used at the University of
Southampton's Highfield Campus in the academic year 2009/2010 for
teaching about 85 Aerospace engineers, and has been used every year
since for growing student numbers, reaching 425 students in
2014/2015.  The Southampton deployment now additionally serves another
cohort of students who study at the University of Southampton Malaysia
Campus (USMC) and there is a further deployment at the Indian
Institute of Technology (IIT) Mandi and Madras campuses, where the
system has been integrated with the Moodle learning management system,
as described in~\secref{sec:iit}.

The testing system has also been used in a number of smaller courses
at Southampton, typically of approximately 20 students, such as one-week
intensive Python programming courses offered to PhD students. It also
serves Southampton's courses in advanced computational
methods where around 100 students have submitted assignments in C, as
described in~\secref{sec:oth-lang}.

\subsection{Case study: Introduction to Computing}

In this section, we present and discuss experience and pertinent statistics from the
production usage of the system in teaching our first-year computing
course, in which programming is a key component. In 2014/15, there
were about 425 students in their first semester of studying
Acoustic Engineering, Aerospace Engineering, Mechanical Engineering,
and Ship Science.

\subsubsection{Course structure}\label{sec:computing1415-outline}

The course is delivered through weekly lectures
(Sec.~\ref{sec:lectures}) and weekly self-paced student exercise
(Sec.~\ref{sec:comp-labor}) with a completion deadline a day before
the next lecture takes place (to allow the lecturer to sight
submissions and provide generic feedback in the lecture the next
day). Students are offered a 90 minute slot (which is called
``computing laboratory'' in Southampton) in which they can carry out
the exercises, and teaching staff are available to provide
help. Students are allowed and able to start the exercise before that
laboratory session, and use the submission and testing system anytime
before, during and after that 90 minute slot.

\begin{figure}
\centerline{\includegraphics[width=0.8\columnwidth]{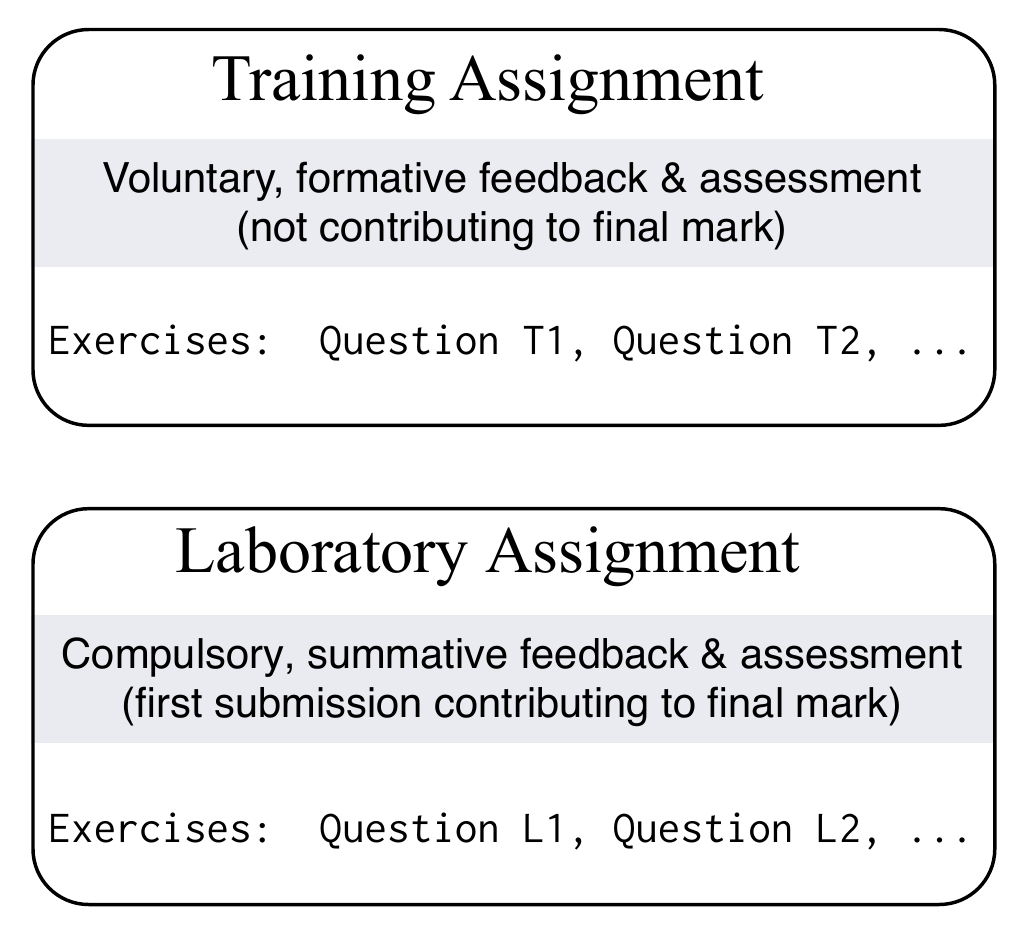}}
\caption{\label{fig:laboratory-assessment-structure}Overview of the
  structure of the weekly computer laboratory session: A voluntary set
  of training exercises is offered to the students as a ``training``
  assignment on which they receive feedback and a mark, followed by a
  compulsory set of exercises in the same topic area as the
  ``laboratory`` assignment which is marked and contributes to each
  student's final mark for the course. Automatic feedback is provided
  for both assignments and repeat submissions are invited.}
\end{figure}

Each weekly exercise is split into two assignments: a set of
``training`` exercises and a set of assessed ``laboratory``
exercises. This is summarised in
Fig.~\ref{fig:laboratory-assessment-structure}.

The training assignment is checked for correctness and marked using the automatic system, but
whilst we record the results and feed back to the students, they do
not influence the students' grades for the course. Training exercises
are voluntary but the students are encouraged to complete them in order to practice the
skills they are currently learning and prepare for the following
assessed exercise which tests broadly similar skills.

Students can
repeatedly re-submit their (modified) code for example until they have
removed all errors from the code. Or they may wish to submit different
implementations to get feedback on those.

The assessed laboratory assignment is the second part of each week's
exercises. For these, the students attempt to develop a solution as
perfect as possible before submitting this by email to the testing
system. This ``laboratory`` submission is assessed and marks and
feedback are provided to the student. These marks are recorded as the
student's mark for that week's exercises, and contribute to the final
course mark. The student is allowed (and encouraged) to submit further
solutions, which will be assessed and feedback provided,
but it is the first submission that is recorded as the student's mark
for that laboratory.

The main assessment of the course is done through a programming exam
at the end of the semester
in which students write code on a computer in a 90 minute session,
without Internet access but having an editor and Python interpreter to
test the code they write. Each weekly assignment contributes of the
order of one percent to the final mark, i.e. 10\% overall for a
10 week course. Each laboratory session can be seen as a training
opportunity for the exam as the format and expectations are similar.

\subsubsection{Student behaviour: exploiting learning opportunities
from multiple submissions}
\label{sec:student-behaviour:exploiting-learning-opportunities-from-multiple-submissions}

\begin{figure}
  \centering
  \begin{subfigure}[b]{0.49\textwidth}
    \includegraphics[width=\textwidth]{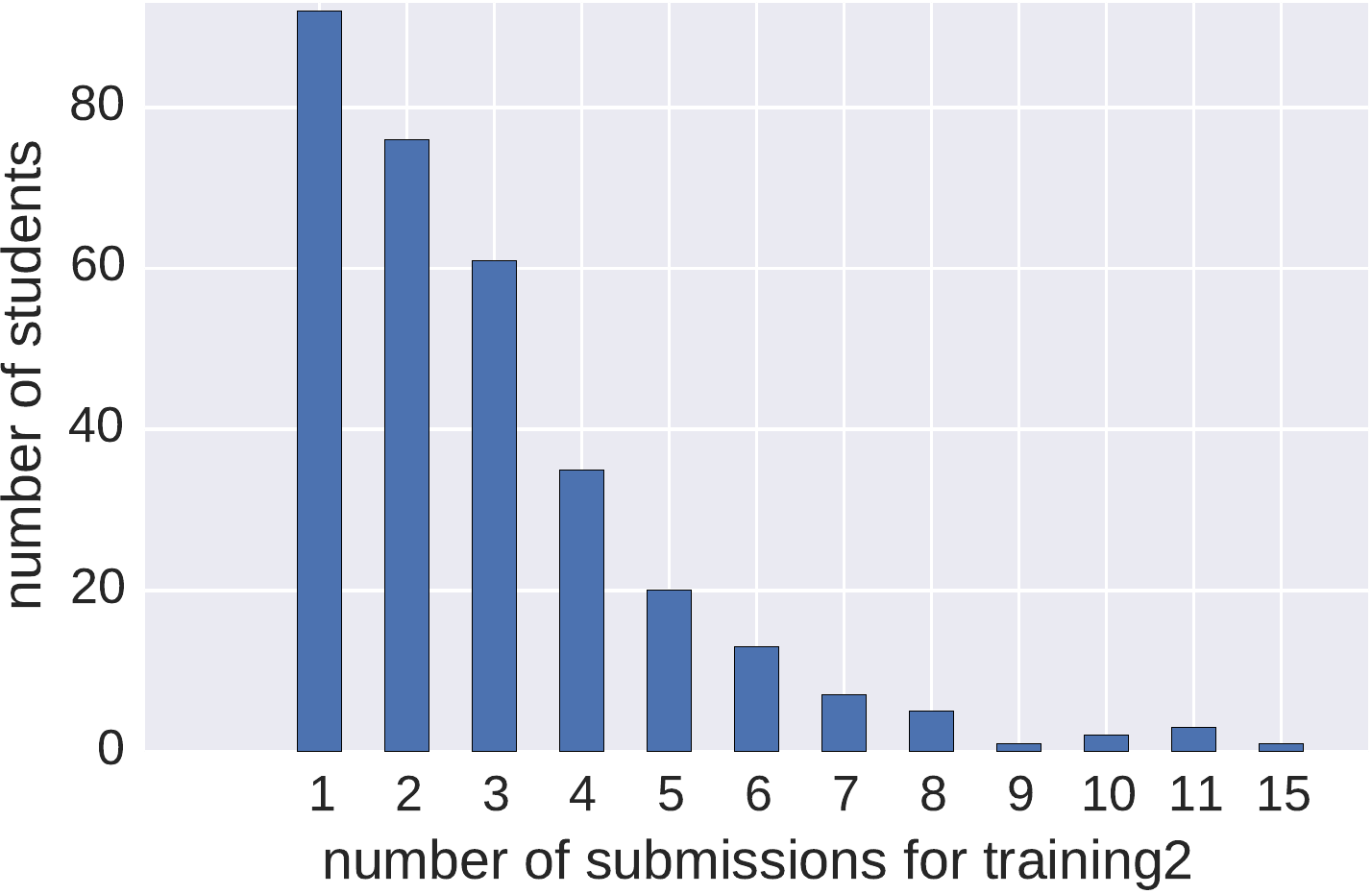}
    \caption{training2\label{fig:tr2-histo}}
  \end{subfigure}
  \begin{subfigure}[b]{0.49\textwidth}
    \includegraphics[width=\textwidth]{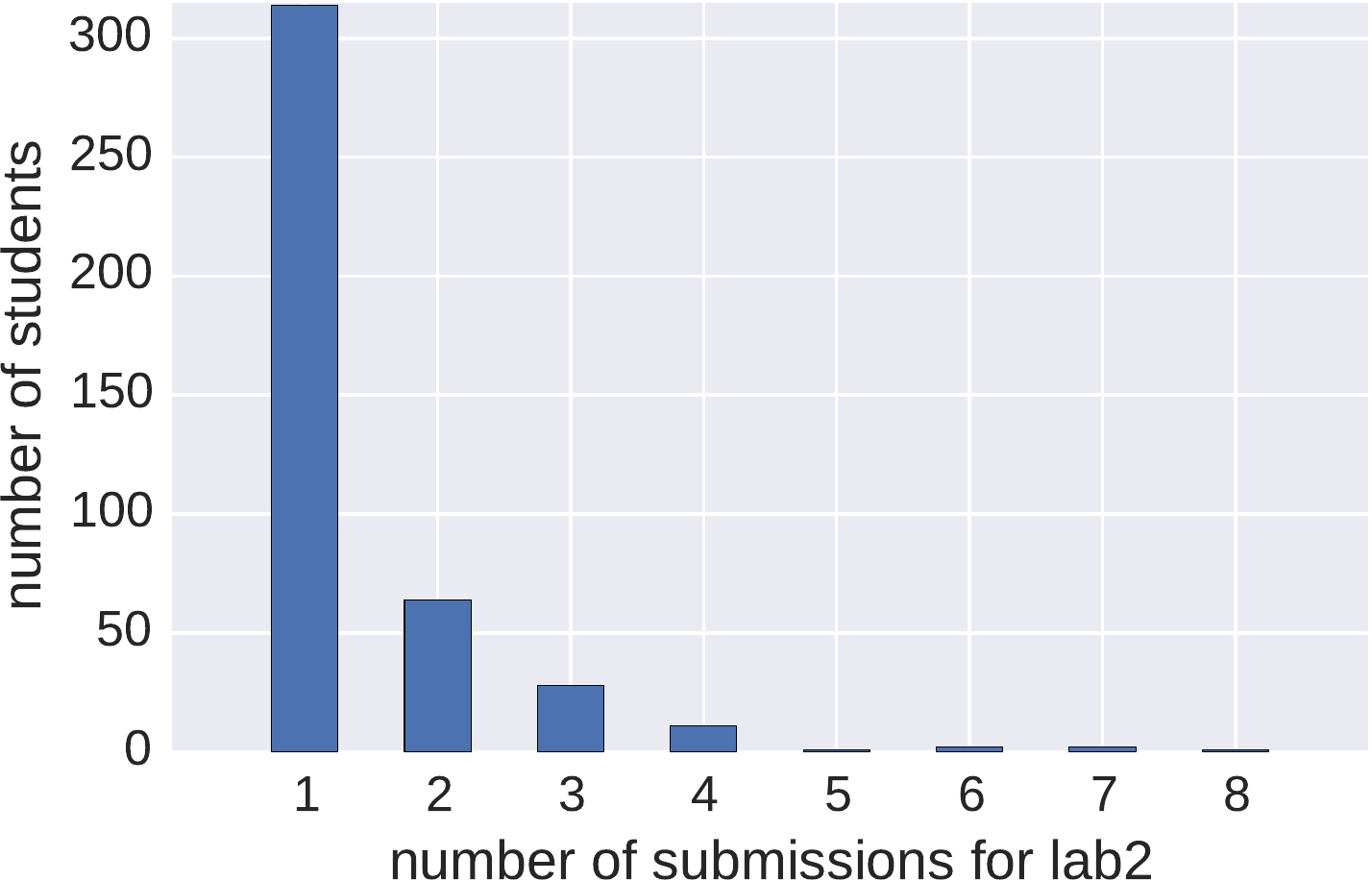}
    \caption{lab2\label{fig:lab2-histo}}
  \end{subfigure}
  \caption{Histogram illustrating the distribution of submission
    counts per student for the (a) voluntary training and (b) assessed laboratory
    assignment (see text in
    Sec.~\ref{sec:student-behaviour:exploiting-learning-opportunities-from-multiple-submissions}) \label{fig:l2-t2-histo}}
\end{figure}

In~\figref{fig:tr2-histo}, we illustrate the distribution of
submission counts for ``training 2'', which is the voluntary set of
exercises from week 2 of the course.

The bar labelled 1 with height 92 shows that 92 students have submitted the
training assignment exactly once, the bar labelled 2 shows that 76 students submitted their training assignment exactly twice, and so on.
The sum over all bars is 316 and shows the total number of students participating in this  voluntary training assignment. 87 students submitted four or more times, and several students
submitted 10 or more times.  This illustrates that our concept of
students being free to make step-wise improvements where needed and
rapidly get further feedback has been successfully realised.

We can contrast this to~\figref{fig:lab2-histo}, which shows the same
data for the compulsory laboratory assignment in week 2
(``lab2''). This submission attracts marks which contribute to the
students' overall grades for the course. In this case the students are
advised that while they are free to submit multiple times for further
feedback, only the mark recorded for their first submission will count
towards their score for the course.  For lab 2, 423 students submitted
work, of whom 314 submitted once only.  However, 64 students submitted
a second revised version and a significant minority of 45 students
submitted three or more times to avail themselves of the benefits of
further feedback after revising their submissions, even though the
subsequent submissions do not affect their mark.

Significant numbers of students choose to submit their work for both voluntary
and compulsory assignments repeatedly, demonstrating that the system offers
the students an extended learning opportunity that the conventional cycle of
submitting work once, having it marked once by a human, and moving to the next
exercise, does not provide.

The proportion of students submitting multiple times for the assessed
laboratory assignment (\figref{fig:lab2-histo}) is smaller than for
the training exercise (\figref{fig:tr2-histo}) and likely to highlight
the difference between the students' approaches to formative and
summative assessment. It is also possible that students need more
iterations to learn new concepts in the training assignment before
applying the new knowledge in the laboratory assignment, contributing
to the difference in resubmissions.  The larger number of students
submitting for the assessed assignment (423 $\approx 100\%$) over the
number of students submitting for the training assignment
($316\approx74\%$) shows that the incentive of having a mark
contribute to their overall grade is a powerful one.

\subsubsection{Student behaviour: timing of submissions}
\label{sec:student-behaviour-timing-of-submissions}

\begin{figure*}
  \centering
  \begin{subfigure}[b]{0.85\textwidth}
    \includegraphics[width=\textwidth]{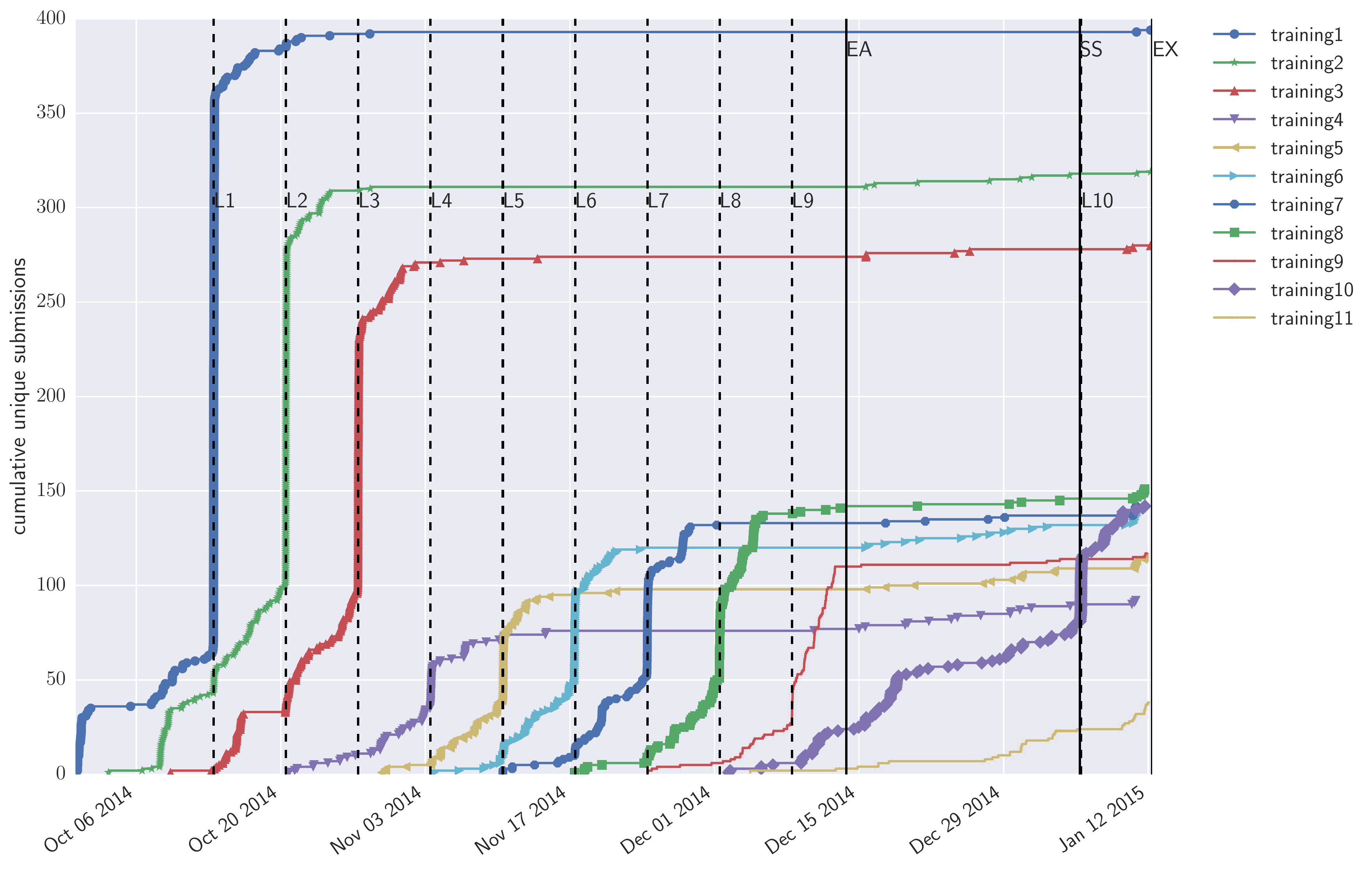}
    \caption{Unique submissions of voluntary training exercises, showing the number of students participating as a function of time.}
    \label{fig:tr-u}
  \end{subfigure}
  \begin{subfigure}[b]{0.85\textwidth}
    \includegraphics[width=\textwidth]{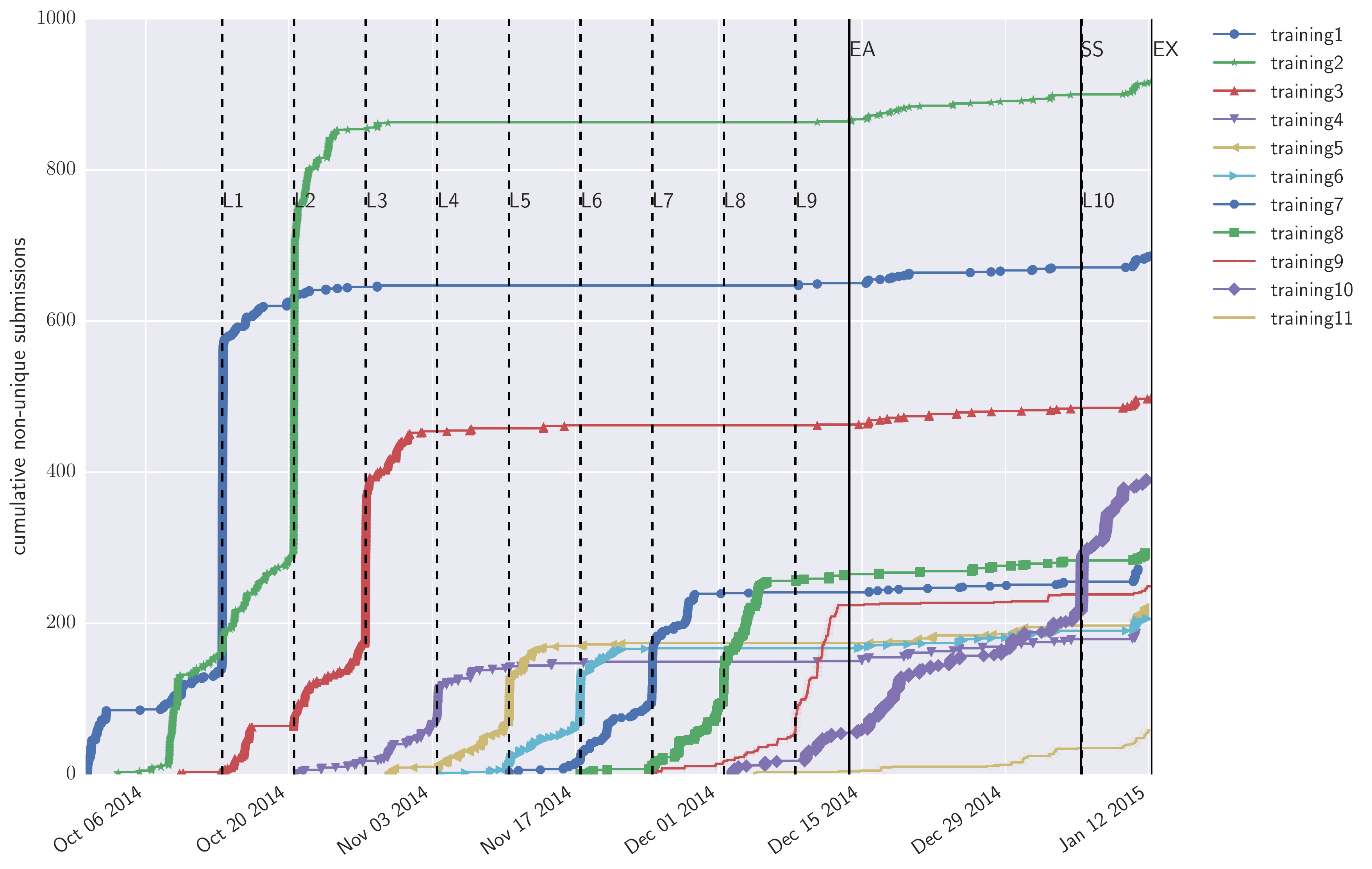}
    \caption{Non-unique submissions of voluntary training exercises, showing the total number of submissions as a function of time.}
    \label{fig:tr-nu}
  \end{subfigure}
  \caption{Submissions of voluntary training assignments as a function
    of time for (a) unique student participation for each assignment,
    (b) total number of submissions for each assignment. Labels
    L1\ldots L10 and associated dashed vertical lines indicate time-tabled computing laboratory sessions 1
    to 10 at Southampton\label{fig:tr-subm-v-time}; EA -- end of
    autumn term; SS -- start of spring term (Christmas break is
    between these dates); EX -- exam. (See
    Sec.~\ref{sec:student-behaviour-timing-of-submissions} for
    details.)}
\end{figure*}

\begin{figure*}
  \centering
  \begin{subfigure}[b]{0.85\textwidth}
    \includegraphics[width=\textwidth]{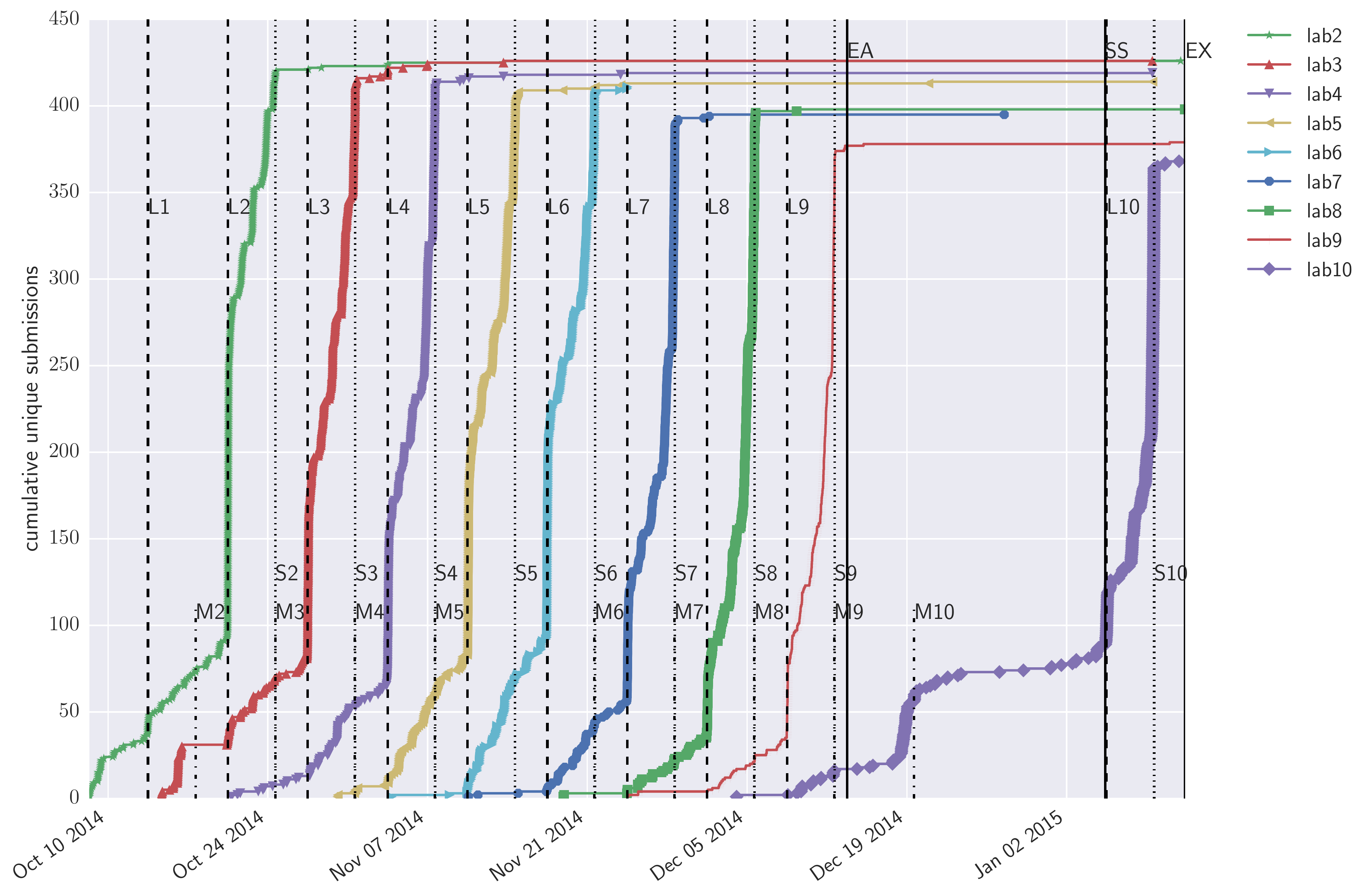}
    \caption{Unique submissions of compulsory assessed laboratory assignments, showing the number of students participating as a function of time.}
    \label{fig:l-u}
  \end{subfigure}
  \quad
  \begin{subfigure}[b]{0.85\textwidth}
    \includegraphics[width=\textwidth]{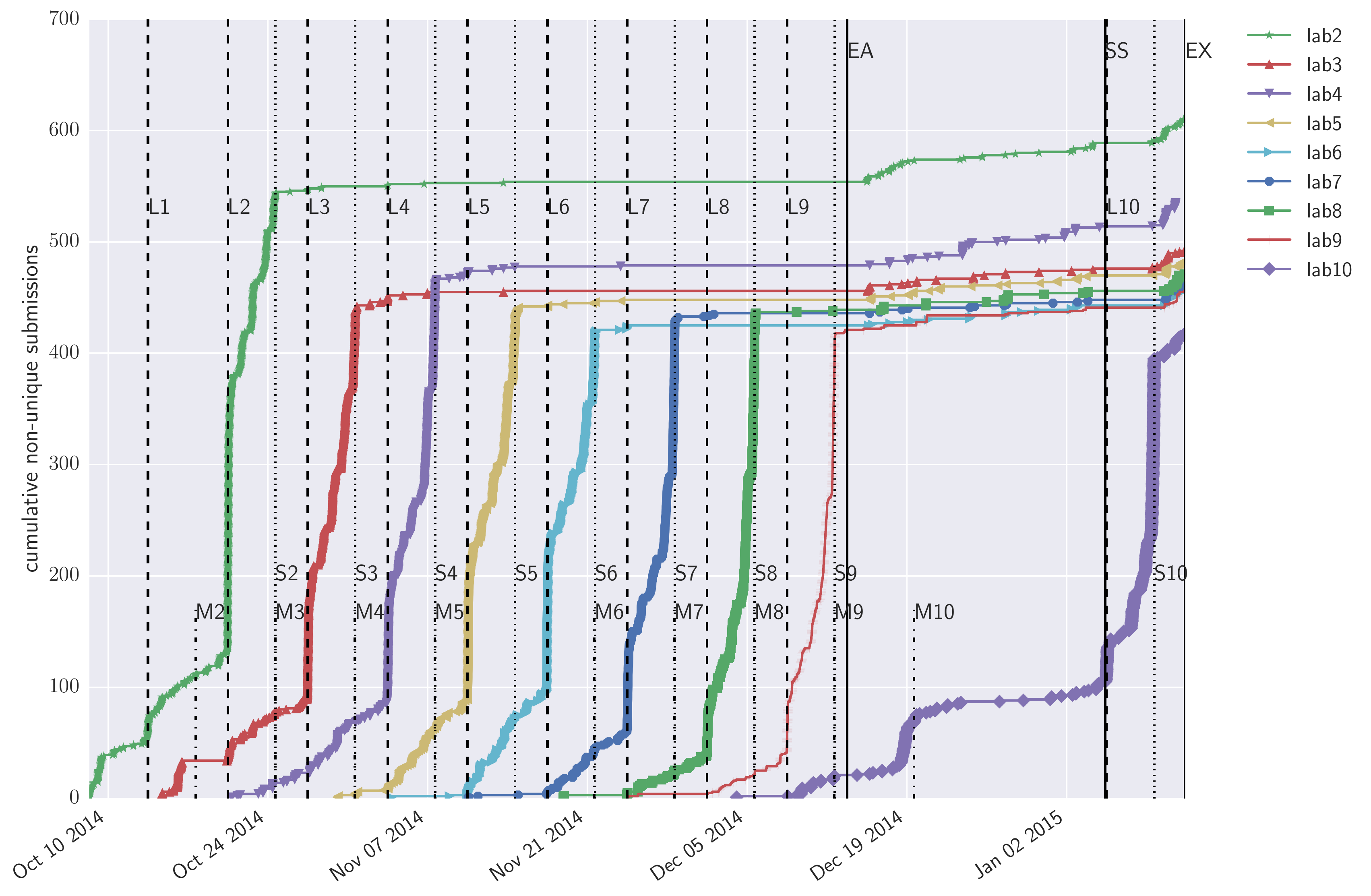}
    \caption{Non-unique submissions of compulsory assessed laboratory assignments, showing the total number of submissions as a function of time.}
    \label{fig:l-nu}
  \end{subfigure}
  \caption{Submissions of compulsory and assessed laboratory assignments as a
    function of time for (a) unique student participation for each
    assignment, (b) total number of submissions for each
    assignment. Labels as in Fig.~\ref{fig:tr-subm-v-time}. Additional
    labels S2 to S10 (vertical dashed lines) indicate submission deadlines in Southampton; M2
    to M10 (vertical dash-dotted lines in lower part of plot) show submission deadlines for students in
    Malaysia. \label{fig:labs-subm-v-time}}
\end{figure*}

In~\figref{fig:tr-subm-v-time} we show the submission timelines for
all the voluntary ``training'' assignments that the students were
offered every week. There are ten such scheduled assignments in total,
and for each a line is shown.  The assignments may be identified by
their chronological sequence, as discussed in the following
paragraphs.

In~\figref{fig:labs-subm-v-time}, we show the same data but for the
compulsory and assessed laboratory assignments
(see Fig.~\ref{fig:laboratory-assessment-structure} and Sec.~\ref{sec:computing1415-outline} for a detailed explanation of the ``training'' and ``laboratory'' assignments).

Plot a) in~\figref{fig:tr-subm-v-time} and a) in
\figref{fig:labs-subm-v-time} show the ``unique'' student submissions
counts for every exercise. With unique, we mean that only the first
submission that any individual student makes for a given assignment is
counted in the graph. On the contrary, subplots b) in
\figref{fig:tr-subm-v-time} and b) in \figref{fig:labs-subm-v-time}
show the ``non-unique'' submissions that include every submission
made, even repeat submissions from any particular student for the same
assignment.

The unique plots allow us to gauge the total number of students
submitting work to a given assignment (as a function of time), and the
non-unique plots allow us to see the total number of submissions made by
the entirety of the student body together.

We discuss the labels and annotations in~\figref{fig:tr-subm-v-time}
first, but they apply similarly to \figref{fig:labs-subm-v-time}. The
dashed vertical lines represent time-tabled computing laboratory
sessions lasting 90 minutes where the students are invited to carry
out the voluntary training and assessed laboratory assignment for that
week in the presence of and with support from teaching staff. These
time-tabled sessions, in which every student has a computer available
to write their code, are labelled L1 to L10 in the figures.

The coloured symbols which are connected by straight lines count the
number of submissions. In~\figref{fig:tr-subm-v-time} a), the
``training 1'' assignment submissions are shown in blue, the next
week's ``training 2`` assignment submissions are shown in green,
etc. There are ten scheduled laboratory sessions, and ten associated
voluntary training assignments. There is one additional assignment at
the end of the course which is offered to help revision for the exam,
shown on the very right in~\figref{fig:tr-subm-v-time} a) and b) in
yellow without symbols.

\figref{fig:labs-subm-v-time} shows submission counts for the
compulsory assessed laboratory exercises. There were 9 such
assessed assignments, starting in the second week of the course:
while there is a voluntary training assignment in week 1, there is no assessed
assignment laboratory assignment to give the students some time to familiarise themselves with the teaching material and submission system. From week 2 onward, there is one voluntary training assignment and one assessed assignment every week up to and including
week 10. The students were given submission deadlines for the
assessed laboratory assignments, and these deadlines for Southampton
students are shown in the plots as dotted vertical lines labelled S2
to S10.

The course was delivered simultaneously at the University of
Southampton (UoS) Highfield Campus in the United Kingdom -- where
about 400 students were taught -- and the University of Southampton
Malaysia Campus (USMC) in Malaysia -- where a smaller group of about
25 students was taught. While following the same lecture material and
assignments, these two campuses, due to different local arrangements
and time zones, taught the course to different schedules, and the
effect of this division is visible in all of the figures. The Malaysia
students have different deadlines from the Southampton students, and
these are shown as shorter vertical dash-dotted lines, labeled M2 to
to M10, towards the bottom of each plot in
\figref{fig:labs-subm-v-time}. The deadlines of students in Malaysia (M) and Southampton (S) follow local holidays and other constraints, although they often fully coincide (S6 to S9), or are delayed by one week (S2 to S5).

We now discuss the actual data presented, starting with the voluntary
training assignment submissions
in~\figref{fig:tr-subm-v-time}. Looking at~\figref{fig:tr-u} we see
that the first training exercise had the largest number of submissions
of any of the training exercises.  About 300 of these submissions
occurred during the first hands-on taught session L1, reflecting a
large number of students who followed the recommended learning
procedure of completing the voluntary exercises and doing so during
the computing laboratory session in the presence of teaching staff,
and who had sufficient resource and instruction available to do so.

The corresponding burst of submissions during the computing laboratory sessions L2 and
L3 has decreased to about 175 and 150 submissions, respectively. The
total number of students participating in these voluntary assignments
in the first three weeks decreases from about 400 in week 1, to about
310 and about 270 in weeks 2 and 3, respectively. The total number of unique
submissions reaches its minimum of about 80 in week 4 (the purple data
set associated with hands-on session L4), and then starts to increase
again for the remainder of the course.

We see in~\figref{fig:labs-subm-v-time} a) that the compulsory
submissions remain high, so that this drop in the voluntary
submissions is no reason for concern, and may reflect that students
understand the learning methods, and options for learning activities
that suit their own preferences and strengths. The data may also
suggest an opportunity to make the assignments slightly more
challenging as students seem to feel very confident in tackling them.

In addition to the burst of submissions during the time-tabled
sessions L1 to L10, we also note a significant number of submissions
both before and after these sessions in~\figref{fig:tr-subm-v-time} a)
and b), reemphasising the flexibility that the system affords students
as to where and when they submit their work. Anecdotal evidence,
written feedback from the students (Sec.~\ref{sec:feedback-from-students}) combined with the submission data suggests that some students will do
the exercises as soon as they become available, and others prefer to do
this during the weekend or evening hours. Many students see the offered
computing laboratory sessions as an opportunity to seek support which
they make use of if they feel this will benefit their learning.

Figure~\ref{fig:tr-subm-v-time}a shows that as the examination date
(labelled as E at the right-hand side of the graphs) approached, a
relatively small number of students started to submit solutions to the
training exercises they had not submitted before as part of their revision
and exam preparation.. The same tendency is visible in~\figref{fig:tr-subm-v-time}b with a slightly larger increase due to repeat submissions that cannot be seen in the graph in Fig.~\ref{fig:tr-subm-v-time}a.

\medskip

We now discuss~\figref{fig:labs-subm-v-time} which shows the same type
of data as~\figref{fig:tr-subm-v-time} but for the compulsory assessed
laboratory assignments rather than voluntary training assignments. The
most notable difference is that the total number of submissions
remained high for all the assignments, reflecting that these
assignments are not voluntary and do contribute to the final course
mark. There is a slow decline of submissions present (from about 425
to 375 during the course, corresponding to approximately 10\%) which
is not unexpected and includes students leaving their degree programme
studies altogether, suspending on health reasons, etc.

The vast majority of first submissions for the compulsory laboratory
assignments, which contribute to the overall course marks, occur in
advance of the deadline, as illustrated in~\figref{fig:l-u} where 
the deadlines are shown as vertical dotted lines.

The assessed assignment timings in~\figref{fig:l-u} show that submissions
take place in different phases. The trend is visible in all the
lines, but most clearly where the submission deadlines in Malaysia coincide
with those in Southampton, i.e. laboratory sessions 6, 7, 8 and 9: the first
submissions are received after the assignments have been published, and then a
steady stream of submissions comes in, leading to an approximately straight diagonal line in~\figref{fig:l-u}. The second set of submissions is received during the
associated laboratory session (shown as dashed line) where many students
complete the work in the timetabled session. Following that, there is again a steady stream of
submissions up to the actual deadline (shown as dotted line) where
submissions accumulate. Very few (first) submissions are received after the
deadline. The submissions in the second phase can be used to estimate student attendance in the laboratory sessions (see discussion in \ref{sec:large-classes}).

The
University of Southampton Christmas break is also apparent, a period
during which there are few new unique submissions
(Fig.~\ref{fig:labs-subm-v-time}a), but slightly more new non-unique
submissions (Fig.~\ref{fig:labs-subm-v-time}b) from students revising
over the holiday and re-submitting assignments they had submitted
before. It is reasonable to assume that they have re-written the code
as an exam preparation exercise.

Trends seen in the voluntary submission data
in~\figref{fig:tr-subm-v-time}, such as a notable rise in the
non-unique (i.e. repeat) submissions across all assignments in the
days leading up to the exam, are also evident in
~\figref{fig:labs-subm-v-time}.

\subsection{Feedback from students}\label{sec:feedback-from-students}

While overall ratings of our courses using the automatic testing and
feedback system are very good, it is hard to distinguish the effect of
the testing system from that of, for example, an enthusiastic team of
teachers, that would also achieve good ratings when using more
conventional assessment and feedback methods.

We invited feedback explicitly on the automatic feedback system
asking for voluntary provision of (i) reasons why students liked the
system and (ii) reasons why students disliked the system. The replies
are not homogeneous enough to compile statistical summaries, but we
provide a representative selection of comments we have received below.

\subsubsection{I like the testing system because\dots}
The following items of feedback were given by the students when
offered to complete the sentence ``\emph{I like the testing system
because\dots}'' as part of the course evaluation:

{\em
\begin{enumerate}
\item because we can get quick feedback
\item it is very quick
\item it provides a quick response
\item immediate effect
\item quick response
\item it gives very quick feedback on whether code has the desired effect
\item it provides speedy feedback, even if working at home in the
  evening \label{item:work-in-the-evenings}
\item it worked and you could submit and re-submit at your own
  pace \label{item:like-resubmitting-possible-1}
\item I like the introduction to the idea of automated unit
  testing. \label{item:like-unittesting}
\item concise, straight to the point, no mess, no fuss. "Got an error?
  Here's where it is. FIX IT!" \label{item:like-see-where-wrong0}
\item it was easy to read output to find bugs in
  programs \label{item:like-see-where-wrong1}
\item you can see where you went
  wrong \label{item:like-see-where-wrong2}
\item very informative, quick response
\item it reassures me quickly about what I do \label{item:like-re-assure}
\item it gave quick feedback and allowed for quick reassessment once
  changes were
  made \label{item:like-refactoring} \label{item:like-resubmitting-possible-2}
\item feedback on quality of the code \label{item:clean-code}
\item it is fast and easy to use
\item it indicates where the errors are and we can submit our work as
  many times as we want \label{item:like-see-where-wrong3}
\item it is quick and automatic
\item it is automated and impartial \label{item:like-objective}
\item gives quick feedback, for training lets you test things quickly
\item it saves time and can give feedback very quickly. The
  re-submission of training exercises is very
  useful. \label{item:like-resubmitting-possible-3}
\end{enumerate}
}
We briefly summarise and discuss these points: the most frequent
student feedback is on the immediate feedback that the system
provides. Some student comments mention explicitly the usefulness of
the system's feedback which allows to identify the errors they have
made more easily (items \ref{item:like-see-where-wrong0},
\ref{item:like-see-where-wrong1}, \ref{item:like-see-where-wrong2},
\ref{item:like-see-where-wrong3}). In addition to these generic
endorsements, some students mention explicitly advantages of the
test-driven development such as re-assurance regarding correctness of
code (item \ref{item:like-re-assure}), quick feedback on refactoring
(\ref{item:like-refactoring}), the indirect introduction of unit tests
through the system (\ref{item:like-unittesting}), and help in writing
clean code (\ref{item:clean-code}). It is worth noting that Agile
methods and test-driven development have not been introduced to the
students at the time where they have provided the above
feedback. Further student feedback welcomes the ability to re-submit
code repeatedly (items \ref{item:like-resubmitting-possible-1},
\ref{item:like-resubmitting-possible-2},
\ref{item:like-resubmitting-possible-3}) and the flexibility to do so
at any time (\ref{item:work-in-the-evenings}). Interestingly, one
student mentions the objectiveness of the system
(\ref{item:like-objective}) -- presumably this comment is based on
experience with assessment systems where a set of markers manually
assess submissions which naturally display some variety in rigour and
the application of marking guidelines.

\subsubsection{I dislike the testing system because\dots}

The following items of feedback were given by the students when
offered to complete the sentence ``\emph{I dislike the testing system
because\dots}'' as part of the course evaluation:
{\em
\begin{enumerate}
\item error messages not easy to
  understand \label{item:dislike-not-easy-to-understand-beginning}
\item it takes some time to understand how to interpret it
\item sometimes difficult to understand what was
  wrong \label{item:dislike-not-easy-to-understand-end}
\item it complains (gives failures) for picky reasons like wrong
  function names and missing docstrings. That's not a complaint, it is
  only a machine. \label{item:dislike-strictness-beginning}
\item it is a bit unforgiving
\item it is extremely [strict] about PEP~8 \label{item:dislike-strictness-3}
\item tiny errors in functions would result in complete failure of
  test. \label{item:dislike-strictness-end}
\end{enumerate}
}
Several comments (items
\ref{item:dislike-not-easy-to-understand-beginning} to
\ref{item:dislike-not-easy-to-understand-end}) state that the feedback
from the automatic testing system is hard to understand. This refers
to test-failure reports such as shown in
Listing~\ref{lst:wrong-res}. Indeed, the learning curve at the
beginning of the course is quite high: the first 90 minute lecture
introduces Python, Hello World and functions, and demonstrates
feedback from the testing system to prepare students for their
self-paced exercises and the automatic feedback they will
receive. However, a systematic explanation of the \texttt{assert}
statements, \texttt{True} and \texttt{False} values, and exceptions
takes only place after the students have used the testing system
repeatedly. The reading of error messages is of course a key skill
(and the importance of this is often underestimated by these
non-computer science students), and we like to think that the early
introduction of error messages from the automatic testing is overall
quite useful. In practice, most students use the hands-on computing
laboratory sessions to learn and understand the error messages with
the help of teaching staff before these are covered in greater detail
in the lectures. See also
Sec.~\ref{sec:quality-of-feedback-provision}.

A second set of comments relates to the harshness and unforgiving
nature of the automatic tests (items
\ref{item:dislike-strictness-beginning} to
\ref{item:dislike-strictness-end}). Item
\ref{item:dislike-strictness-end} refers to the assessment method of
not awarding any points for one of multiple exercises that form an
assignment if there is any mistake in the exercise, and is a criticism
regarding the assessment as part of the learning process.

For items \ref{item:dislike-strictness-beginning} to
\ref{item:dislike-strictness-3} it is not clear whether these
statements relate to the feedback on the code or the assessment. If
the comments relate to the code, then they reflect a lack of
understanding (and thus a shortcoming in our teaching) of the
importance of documenting code and the importance of getting
everything right in developing software (and not just approximately
right).

\subsubsection{Generic comments}
The following comments on the feedback system were provided by
students unprompted, i.e. as part of generic feedback on the course,
and are in-line with the more detailed points made above:
{\em
\begin{enumerate}
\item Fantastic real-time feedback with online submission of exercises.
\item Loved the online submission.
\item Really like the online submission system with very quick feedback.
\item Description in the feedback by automated system can be unclear.
\item Instant feedback on lab and training exercises was welcome.
\item Autotesting feature is VERY useful! Keep it and extend it!
\item The automatic feedback is fairly useful, once you have worked
  out how to understand it.
\end{enumerate}
} 
In the context of enthusiastic endorsements of the testing system, we
like to add our subjective observation from teaching the course that
many students seem to regard the process of making their code pass the
automatic tests as a challenge or game which they play against the
testing system, and that they experience great enjoyment when they
pass all the tests -- be it in the first or a repeat submission (see
also~\secref{sec:iterative-tests}). As students like this game, they
very much look forward to being able to start the next set of
exercises which is a great motivation to actively follow and
participate in all the teaching activities.

\subsection{Issues}

During the years of using the automatic testing system, we have
experienced a number of issues which are unique to the automated
method of assessment described here. We summarise them and our
response to each challenge below.

\subsubsection{Submissions including syntax errors}
\label{sec:Submissions-including-syntax-errors}

When a student submits a file containing a syntax error, our testing
code (here driven by the \verb+py.test+ framework) is unable to import
the submission, and therefore testing cannot commence. Technically,
such a submission is not a valid Python program (because it contains
at least one syntax error). We ask the students to always test their
work thoroughly before submitting, which should detect syntax errors
first, and such submissions should not occur.

However, in practice, and given the large number of submissions (about
20 assignments per student, and currently 500 students per year),
occasionally students will either forego the testing to save time, or
will inadvertently introduce syntax errors such as additional spaces
or indentation between checking their work and submitting it.  From a
purely technical point of view, the system is able to recognise this
situation when it arises and we could state that any such submission
is incorrect, and therefore assign a zero mark.  However, these
submissions may represent significant effort and contain a lot of
valid code (for multiple exercises submitted in one file), so we have
adopted a policy of allowing re-submission in such a scenario: if a
syntax error is detected on import of the submission, the student is
automatically informed about this, and re-submission is invited.

\subsubsection{Submissions in undeclared non-ASCII character encoding}

We noticed an increasing trend, especially among international
students, for submitting files in 8-bit character sets other than
ASCII.  Such files are accepted by the Python 2 interpreter so long as
the encoding is declared in the first lines of the file according to
the PEP263~\cite{PEP263}; but many of the students who were using
non-ASCII characters were not describing their encodings at all.  Our
first response was to update our system to check for this situation,
and upon discovering it, to send an automated email to the student
concerned with a suggestion that they declare their encoding and
re-submit.  More recently, we have began recommending the use of the
Spyder \cite{Spyder} environment, whose default behaviour is to
annotate the encoding of the file in question in a PEP 263--compliant
manner.  This has now virtually eliminated the occurrence of character
encoding issues.  For the few cases where these still arise, the
automatic suggestion email, and (if required) personal support in
scheduled laboratory and help sessions enables the students to
understand and overcome the issue.

\subsubsection{PEP~8 style checker issues}

As described in~\secref{sec:tests}, we take advantage of the
\verb+pep8+ utility \cite{pep8tool} to assess the conformance of the
students' submissions against the style recommendations of PEP~8.

Students find following style guidelines a lot harder than adapting to
hard syntactic and semantic requirements of the programming language
as they can solve the given exercises so that their code exhibits
correct functional behaviour while not necessarily following the style
guidelines. In our experience, it is critical to help students to
adapt their own style habits to recommended guidelines, for example
through tools that flag up non-confirming constructs immediately while
editing code. One such freely available tool for Python is the Spyder
\cite{Spyder} development environment, for which PEP~8-compatibility
highlighting can be activated \cite{spyder-tutorial}. By encouraging
all students to use this environment -- at university machines and in
installations of the software on their own machines for which we
provide recommendations \cite{anaconda-installation-2015} -- we find
that they generally pick up the PEP~8 guidelines quickly. As with so
many things, if introduced early on, they soon embrace the approach
and use it without additional effort in the future. Consequently, we
penalise submissions that are not PEP~8 compliant from the second week
onward.

One issue that arises with integrating PEP~8 guidelines into the
assessment is that different software release versions of the
\texttt{pep8} tool may yield different numbers of warnings; this is
partly due to changes in the view on what represents good coding style over time
 and partly due to bugs being fixed in the
\texttt{pep8} tool itself. This can result in unexpected warnings from
the PEP~8-related tests. As a practical measure, we ensured that we
are using the latest version of the \verb+pep8+ checking tool, and
have elected to omit those tests that are treated differently by other
recent versions. The student body will generally report any such
deviations between the PEP~8 behaviour on their own computer and the
testing system, and help in identifying any potential problems here.

\subsection{Integration with Moodle\label{sec:iit}}

Moodle (Modular Object-Oriented Dynamic Learning Environment) \cite{Moodle} is a
widely-used open source learning management system which can be
used to deliver course content and host online learning activities. It
is designed to support both teaching and learning activities. The Indian
Institute of Technology (IIT)
Mandi and IIT Madras use Moodle to manage the courses at the institute
level.  When running a course, instructors can add resources and
activities for their students to complete, \emph{e.g.} a simple page
with downloadable documents or submission of the assignments by
prescribed time and date.

It was envisaged that integrating the automatic feedback provision
system with Moodle would simplify the use of the automatic feedback
system for IIT instructors and students, by allowing to submit and retrieve feedback through the Moodle interface that they use routinely already instead of using email, thus
replacing the incoming queue process (Fig.~\ref{fig:flowdiagram-incoming}). Outgoing messages to administrators are still emailed using the outgoing email queue (Fig.~\ref{fig:flowdiagram-outgoing}). The testing process queue (Fig.~\ref{fig:flowdiagram-testing}) is used as in the Southampton deployment that is described in the main part of this paper.

In integrating the assessment system with the IIT
Moodle deployment, we have used the Sharable Content Object Reference
Model, SCORM, which is a set of technical standards for e-learning
software products.
The user front end
is provided through the browser-based Moodle User Interface, while scripts at the back end
make the connection to the automatic assessment system. The results
are then fetched from the system and made visible to the student and
the instructor. Using Moodle also helps the IIT to leverage the
security that is already a part of the SCORM protocols.

The implementation at IIT is via a Moodle plugin designed such that,
when a student submits an assignment, the plugin collects the global
file ID of the submission and creates a copy of the file outside the
Moodle stack. The plugin then invokes a Python script through
\verb+exec()+, transferring the location of file and file ID to the
script.  This Python script then acts as a user of the automatic
feedback and assessment system, and directly enqueues the file for
processing. The job ID inside automatic assessment system engine is
returned to the Moodle plugin which maintains a database mapping job
IDs to file IDs. After the file is processed by the automatic
assessment system, the results are saved as files that are named after
the (unique) job ID.  When students access their results through
Moodle, the relevant job IDs are retrieved from the database, allowing
the corresponding results file to be opened, converted to HTML and
published in a new page.

\subsection{Testing of other languages\label{sec:oth-lang}}

There should be no conceptual barriers for using the automatic testing and
feedback system for testing of code written in other languages that provide
unit testing frameworks. In particular JUnit \cite{JUnit} for Java could be
used instead of \texttt{py.test} for Java programming courses. In this case,
the execution of the actual tests (and the writing of the tests to run) would
need to be done in Java, but the remaining framework implemented in Python
could remain (mostly) unchanged, providing the student submissions handling and
receipts, separation of testing jobs, a limited-privilege, limited-resource
runtime environment, maintaining a database of results, and automatically
emailing students their feedback.

\medskip

As part of our education programme in computational science \cite{Fangohr:computational-science-essential-tools}, we we are
interested in testing C code that students write in our advanced
computational methods courses in which they are introduced to C programming. Students
learn in particular how to combine C and Python code to benefit from Python's
effectiveness as a high level language but achieving high execution
performance by implementing performance critical sections in~C.

We are exploring a
set of light weight options towards automating the testing of
the C code within the given framework and our education setting:
\begin{enumerate}
\item Firstly, we compile the submitted C code using \verb+gcc+,
  capturing and parsing its standard output and standard error to
  capture the number of errors and warnings generated.
\item We then run the generated executable under the same security
  restrictions as we use for Python, capturing its standard output and
  error, and potentially comparing them to known-correct examples.
\item We are also using the \verb+ctypes+ library to make functions
  compiled from students' C code available within Python, so that they
  may be tested with tests defined the same way as for native Python
  code (see~\lstref{lst:testdef}).
\end{enumerate}
The system that we built for testing Python code is modular enough
that the above can be incorporated into the
test work-flow for the courses where it is required. We note that it is now necessary to handle segmentation faults that may arise from calling the student's C code: these may be treated similarly to the cases where resource limits are exceeded in testing Python code, causing the OS to terminate the process; the student's marks may be updated if required, or a re-submission invited, in line with the course leader's chosen educational policy.

\subsection{Pre-marking exams}
As well as assessing routine laboratory assignments, the system is
also used to support exam marking. The format of the exam for our
first year introductory programming course is a 90 minute session
which the students spend at a computer, in a restricted
environment. They are given access to the Spyder Python development
environment to be able to write and run code but have no access to the
Internet, and have to write code to answer exam questions which follow
the format experienced in the weekly assignments. At the end of the
exam, all the students' code files are collected electronically for
assessment. We \emph{pre-test} the exam code files using the marking
system with an appropriate suite of tests, and then distribute
the automatically assigned marks and detailed test results
and the source code to the examiners
for manual marking. This enables the examiners to save significant
amounts of time because it is immediately apparent when students
achieve full marks and, where errors are found, the system's output
assists in swiftly locating them. It also increases objectivity
compared to leaving all the assessment to be done by hand, possibly by a team
of markers who would each have to interpret and apply a mark scheme to
the exam code files.

The system has also been used to receive coursework submissions for a
course leader who decided to exclusively manually assess the work. In
this case, the system was configured simply to receive the submission,
identify the user, store the submission, and log the date and time of
submission of the coursework.

\subsection{Discussion}

In this section, we discuss key aspects of the design, use and
effectiveness of the automatic testing system to support learning of
programming.

\subsubsection{Key benefits of automatic testing}\label{sec:key-benefits}

A key benefit of using the automatic testing system is to reduce the
amount of repeated algorithmic work that needs to be carried out by
teaching staff. In particular establishing the correctness of student
solutions, and providing basic feedback on their code solutions is now
virtually free (once the testing code has been written)
as it can be done automatically.

This allowed us to very significantly increase the number of exercises
that students carry out as part of the course, which helped the
students to more actively engage with the content and resulted in deeper
learning and greater student satisfaction.

The marking system frees teaching staff time that would otherwise have
been devoted to manual marking, and which can now be used to repeat
material where necessary, explain concepts, discuss elegance,
cleanness, readability and effectiveness of code, and suggest
alternative or advanced solution designs to those who are interested,
without having to increase the number of contact hours.

Because of the more effective learning through active self-paced exercises, we
have also been able to increase the breadth and depth of materials in some of
our courses without increasing contact time or student time devoted to
the course.

\subsubsection{Quality of automatic feedback provision}\label{sec:quality-of-feedback-provision}
The quality of the feedback provision involves two main aspects: (i)
the timeliness, and (ii) the usefulness, of the feedback.

The system typically provides feedback to students within 2 to 3
minutes of their submission (inclusive of an email round-trip time on
the order of a couple of minutes).  This speed of feedback provision
allows and encourages students to iteratively improve submissions
where problems are detected, addressing one issue at a time, and
learning from their mistakes each time.

This near-instant feedback is almost as good as one could hope for,
and is a very dramatic improvement on the situation without the system
in place (where the provision of feedback would be within a week of
the deadline, when an academic or demonstrator is available in the
next practical laboratory session).

The usefulness of the feedback is dependent upon the student's ability
to understand it, and this is a skill that takes time and practice to
acquire.  We elected to use the traceback output provided by
\verb+py.test+ in the feedback emails that are sent to students in the
case of a test failure, as per the example in~\lstref{lst:wrong-res}.
The traceback, combined with our helpful comments in the test
definitions, allows a student to understand under precisely which
circumstances their code failed, and also to understand why we are
testing with that particular set of parameters.  Although interpreting
the tracebacks is not a skill that is immediately obvious, especially
to students who have never programmed before, it is a skill that is
usually quickly acquired, and one which all competent programmers
should be well-versed in. We suggest that it is an advantage to encourage students to
develop this ability at an early stage of their learning.
Students at Southampton are well-supported in acquiring these skills, including timetabled weekly laboratories and help sessions staffed by academics
and demonstrators.  Once the students master reading the output, the
usefulness of the feedback is very good: it pinpoints exactly where
the error was found, and provides the rationale for the choice of test
case as well.

A third aspect of the quality of feedback and assessment is
objectivity. Because all of the submissions are tested to the same
criteria, the system also improves the objectivity of our marking
compared to having several people each interpreting the mark scheme
and applying their interpretations to student work.

\subsubsection{Flexible learning opportunities}\label{sec:flexible-learning}

A further enhancement to the student experience is that the system
allows and promotes flexible working.  Feedback is available to
students from anywhere in the world (assuming they have Internet access), at any time of day or night,
rather than being restricted to the locations and hours that laboratory
sessions are scheduled.  This means that the most confident students
are free to work at their own pace and convenience.  Those students
who wish for more guidance and support can avail themselves of the
full resources in the time tabled sessions.  All the students can
repeat training exercises multiple times, dealing with an error at
once, when errors are discovered.  They may also repeat and re-submit
assessed laboratory exercises to gain additional feedback and deeper
understanding, but in line with our policies, this does not change
their recorded marks for assessed work.

\subsubsection{Large classes}
\label{sec:large-classes}

We have found that the assessment system is invaluable as our student
numbers grow between years. Once exercises and didactic testing code
are developed, the automatic testing and feedback provision does not
require additional staff time to process, assess and feedback on
student submissions when student numbers grow from year to
year. Additional teaching staff in the practical sessions are required
to maintain the student-staff ratio, but the automatic system reduces
the overall burden very significantly, and has helped us to deliver
the training in the face of an increase from 85 to 425 students
enrolled in our first-year introduction to computing course.

The flexible learning that the system allows (see
Sect.~\ref{sec:flexible-learning}) holds opportunities for more
efficient space use. In the weekly hands-on computing laboratories, we
currently provide all students their own computer for 90 minutes in
the presence of teaching staff. With large student numbers, depending
on the local facilities, this can become a time tabling and resource challenge.

We know from student attendance behaviour that the first two weeks see
nearly all students attending the hands-on computing laboratories, but
that student attendance in the computing laboratories declines
significantly after week two, as -- for example -- the best students will
often have completed and submitted the exercise before the time-tabled
laboratory session, and some students will only come to the laboratory
session to get help on a particular problem that they could not solve on
their own; needing 15 minutes attendance rather than 90. As a result, it should
be possible to 'overbook' computing laboratory spaces as is common in
the airline industry, for example based on the assumption that only a
fraction of the students will make regular use of the laboratory
sessions in the later weeks. Figure~\ref{fig:l-u} and its discussion shows supporting data of student laboratory attendance. We have not made use of this yet.

\subsubsection{Student satisfaction}

Student feedback on the automatic testing and learning with it has
been overall very positive. We believe that the increased number of
practical exercises is an effective way to educate students to become
better programmers, and it is gratifying for teaching staff to see
students enjoying the learning experience.

\subsubsection{Software design}

Our system design of having multiple loosely coupled processes that
process student submissions with clearly defined sub-tasks,
and pass jobs from one to another through file-system based queues has
provided a robust system, which allowed us to connect it with other
tools, such as for example the Moodle front-end for code submission in
Madras and Mandi.

\section{Summary}
\label{sec:summary}
We have reported on the automatic marking and feedback system that we
developed and deployed for teaching programming to large classes of
undergraduates. We provided statistics from one year of use of our
live system, illustrating that the students took good advantage of the
``iterative refinement'' model that the system was conceived to
support, and that they also benefited from increased flexibility and
choice regarding when they work on, and submit, assignments. The
system has also helped reduce staff time spent on administration and
manual marking duties, so that the available time can be spent more
effectively supporting those students who need this. Attempting to
address some of the shortcomings of other literature in the field as
perceived by a recent review article, we provided copious technical
details of our implementation. With increasing class sizes forecast
for the future, we foresee this system continuing to provide us value
and economy whilst giving students the benefit of prompt, efficient
and impartial feedback. We also envisage further refining the system's
capabilities at assessing submissions in languages other than Python.

\subsection*{Acknowledgements}
This work was supported by the British Council, the Engineering and Physical Sciences Research Council (EPSRC) Doctoral Training grant EP/G03690X/1 and EP/L015382/1. We provide data shown in the figures in the supplementary material.

\bibliography{tetepy-paper-bibliography}{}
\bibliographystyle{IEEEtran}

\end{document}